%% file: sample-sigconf.tex
\documentclass[manuscript,nonacm,screen]{acmart}

\AtBeginDocument{%
  }

\setcopyright{acmlicensed}
\copyrightyear{2026}
\acmYear{2026}
\acmDOI{XXXXXXX.XXXXXXX}
\acmConference[ACM SIGCHI]{Make sure to enter the correct
  conference title from your rights confirmation email}{June 03--05,
  2018}{Woodstock, NY}
\acmISBN{978-1-4503-XXXX-X/2018/06}

\usepackage{array}
\usepackage{float}
\usepackage{graphicx}
\usepackage{subcaption}
\usepackage{booktabs}      % For \toprule, \midrule, \bottomrule
\usepackage{makecell}      % For nicer cell formatting (optional)
\usepackage{xcolor}        % Optional, for color
\usepackage{colortbl}      % Optional, for colored rows/columns
\usepackage[table,xcdraw]{xcolor}
\usepackage{bbding}
\usepackage{wrapfig}
\usepackage[most]{tcolorbox} % in preamble
\usepackage{xcolor}
\usepackage{graphicx}
\usepackage{wrapfig}
\usepackage[most]{tcolorbox}
\tcbuselibrary{skins,breakable}
\usepackage{hyperref}
\usepackage{multirow}

\tcbset{
  myboxstyle/.style={
    colback=green!5!white,   % background color
    colframe=blue!60!black, % border color
    boxrule=0.8pt,          % border thickness
    arc=5mm,                % corner roundness
    left=2mm, right=2mm, top=2mm, bottom=2mm,
    enhanced,
  }
}

\tcbset{ mysidebysidebox/.style={ enhanced, breakable, sidebyside, sidebyside align=center, righthand width=0.4\textwidth, boxrule=2.5pt, arc=5mm, left=2mm, right=2mm, top=2mm, bottom=2mm, } }

\definecolor{G4color}{RGB}{186,223,121}       % Sattes Blau
\definecolor{G4lpcolor}{RGB}{248,214,232}    % Türkisblau
\definecolor{S8ccolor}{RGB}{194,190,214}       % Orange
\definecolor{C8color}{RGB}{242,180,109}       % Dunkelgrün
\definecolor{G8color}{RGB}{147,185,209}    % Violett-Rosa
\definecolor{S8scolor}{RGB}{150,209,198}       % Rot-Orange
\definecolor{V8color}{RGB}{243,137,123}

\newcommand{\MethodBox}[2]{%
  \raisebox{0.5pt}{\textcolor{#1}{\rule{1.2ex}{1.2ex}}}%
  \hspace{0.4ex}\textbf{\textcolor{black}{\large #2}}%
}

\newcommand{\GFour}{\MethodBox{G4color}{\textbf{G4}}}
\newcommand{\GFourLP}{\MethodBox{G4lpcolor}{\textbf{G4~lp}}}
\newcommand{\GEight}{\MethodBox{G8color}{\textbf{G8}}}
\newcommand{\CEight}{\MethodBox{C8color}{\textbf{C8}}}
\newcommand{\TEightC}{\MethodBox{S8ccolor}{\textbf{S8~c}}}
\newcommand{\TEightS}{\MethodBox{S8scolor}{\textbf{S8~s}}}
\newcommand{\VEight}{\MethodBox{V8color}{\textbf{V8}}}

\begin{document}

\title{Evaluating Keyframe Layouts for Visual Known-Item Search in Homogeneous Collections}

\author{Bastian Jäckl}
\email{bastian.jaeckl@uni-konstanz.de}
\orcid{0009-0004-3341-1524}
\affiliation{%
  \institution{University of Konstanz}
  \city{Konstanz}
  \country{Germany}
}

\author{Jiří Kruchina}
\affiliation{%
  \institution{Charles University}
  \city{Prague}
  \country{Czechia}}
\email{kruchi@centrum.cz}
\orcid{0009-0009-3025-5489}

\author{Lucas Joos}
\email{lucas.joos@uni-konstanz.de}
\orcid{0000-0001-7049-5203}
\affiliation{%
  \institution{University of Konstanz}
  \city{Konstanz}
  \country{Germany}
}
\author{Daniel A. Keim}
\email{daniel.keim@uni-konstanz.de}
\orcid{0000-0001-7966-9740}
\affiliation{%
  \institution{University of Konstanz}
  \city{Konstanz}
  \country{Germany}
}
\author{Ladislav Peška}
\orcid{0000-0001-8082-4509}
\affiliation{%
  \institution{Charles University}
  \city{Prague}
  \country{Czechia}}
\email{ladislav.peska@matfyz.cuni.cz}

\author{Jakub Lokoč}
\orcid{0000-0001-8082-4509}
\affiliation{%
  \institution{Charles University}
  \city{Prague}
  \country{Czechia}}
\email{jakub.lokoc@matfyz.cuni.cz}
\renewcommand{\shortauthors}{Trovato et al.}

\begin{abstract}
%Multimodal deep learning networks form the backbone of modern interactive video retrieval systems, enabling semantic ranking of representative keyframes in response to textual queries. 
%Despite these advances, users still need to manually browse ranked keyframe candidates to identify a specific target, a process that can be time-consuming and error-prone.
%The arrangement of candidate keyframes in a search grid highly affects the effectiveness and efficiency for this task, and is still under-explored.
%We present a user study with 49 participants evaluating the browsing performance of seven commonly used 2D keyframe grid layouts on the Visual Known-Item Search task, where participants try to locate a predefined target frame among a set of candidates. 
%In addition to measuring efficiency and accuracy, we relate browsing phenomena, such as overlooks, region skipping, and scrolling behavior, to layout characteristics.
%Our results reveal that a video-grouped design is the most efficient overall, while a four-column rank-preserving grid with a left control panel achieves the highest accuracy. 
%We also provide evidence for the potential of sorted grids, which enable rapid scanning of low-likelihood regions, but down-rank targets into less prominent positions, delaying first arrival times and increasing overlooks. These findings motivate hybrid designs that preserve top-ranked items while sorting or grouping the remainder and offer guidance for manual search in grid-like displays beyond video retrieval.
Multimodal deep-learning models power interactive video retrieval by ranking keyframes in response to textual queries. 
Despite these advances, users must still browse ranked candidates manually to locate a target.
Keyframe arrangement within the search grid highly affects browsing effectiveness and user efficiency, yet remains underexplored. 
We report a study with 49 participants evaluating seven keyframe layouts for the Visual Known-Item Search task.
Beyond efficiency and accuracy, we relate browsing phenomena, such as overlooks, to layout characteristics.
Our results show that a video-grouped layout is the most efficient, while a four-column, rank-preserving grid achieves the highest accuracy. Sorted grids reveal potentials and trade-offs, enabling rapid scanning of uninteresting regions but down-ranking relevant targets to less prominent positions, delaying first arrival times and increasing overlooks.
 These findings motivate hybrid designs that preserve positions of top-ranked items while sorting or grouping the remainder, and offer guidance for searching in grids beyond video retrieval.
\end{abstract}



\begin{teaserfigure}
  \includegraphics[width=\textwidth]{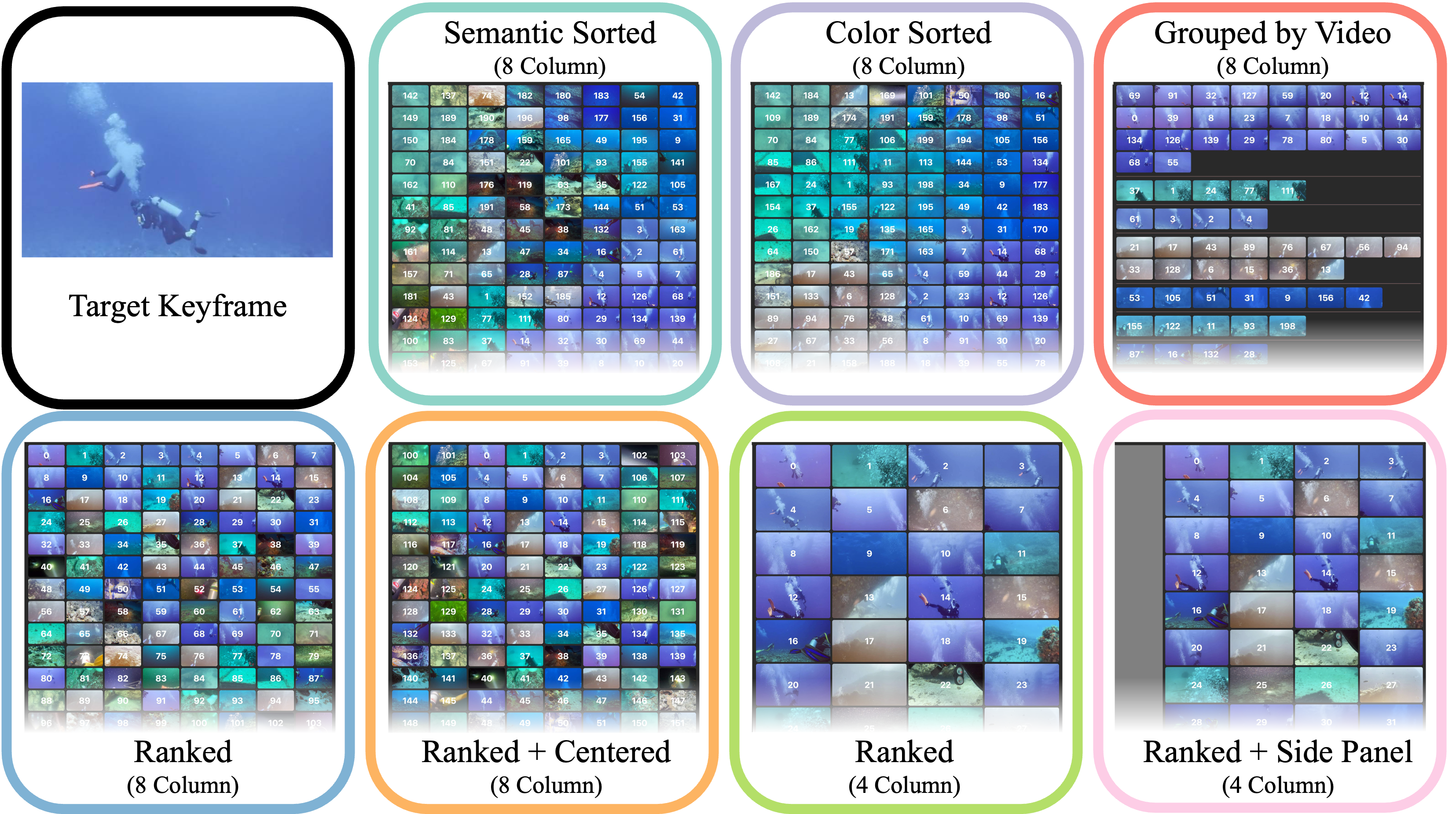}
  \caption{We compare the efficiency and effectiveness of seven keyframe layout strategies for the Visual Known-Item Search task, where the goal is to find a particular target keyframe among a set of candidate keyframes. Each layout presents video content in a distinct way for exploration.}
  \Description{Enjoying the baseball game from the third-base
  seats. Ichiro Suzuki preparing to bat.}
  \label{fig:layout_comparison}
  \Description{This figure illustrates the Visual Known-Item Search task and the seven layouts examined in our work. One box shows the target image (the "known item”). The other boxes show the same set of candidate keyframes arranged in different layouts. Every candidate thumbnail carries a small number indicating its rank. The ranked layouts place items according to rank order. The sorted layouts group visually or semantically similar items. Consequently, rank numbers appear random as the rank does not determine placement. The video-grouped layout clusters frames by their video affiliation. Each video starts a new line, and if a video does not fill a row, the grid contains visible blank spaces. The figure communicates that different layout strategies change both the rank order and neighborhood structure.}
\end{teaserfigure}

\received{20 February 2007}
\received[revised]{12 March 2009}
\received[accepted]{5 June 2009}

\maketitle

\section{Introduction}
\input{Chapter/1-introduction}

\section{Related work}
\label{sec:relatedWork}
\input{Chapter/2-related-work}

\section{User Study Specification}
\label{sec:StudySpecification}
\input{Chapter/3-study-setup}

\section{Results \& Analysis}
\input{Chapter/4-results}

\section{Discussion \& Future Work}
\input{Chapter/5-limitations}

\section{Conclusion}
\input{Chapter/6-conclusion}

\bibliographystyle{ACM-Reference-Format}
\bibliography{references}

\clearpage

\appendix
\input{Chapter/7-appendix}

\end{document}

%% file: Chapter/1-introduction.tex
Searching for specific information in video content is an increasingly common and important task in both everyday life and professional domains.
Whether journalists are looking for a key moment in social media footage~\cite{news_footage}, security analysts are reviewing surveillance videos ~\cite{video_surveilance}, or everyday users are trying to find a particular scene in a personal video collection~\cite{vbs2024_eval_performance}--efficient and accurate video retrieval applications are essential.
In recent years, interactive video retrieval systems have made great strides, primarily driven by advances in multi-modal deep neural networks~\cite{SchallBHJ24, clip_2021} that enable users to issue simple text queries to search for semantic content.
In practice, these retrieval systems operate efficiently on a preprocessed set of representative images, called \emph{keyframes}~\cite{keyframeIntroduction}, rather than entire videos.
Provided with a query, the retrieval system returns a set of best-matching keyframes, which consequently serve as entry points for users to assess and browse this candidate pool. 
Based on this browsing phase, users initiate their subsequent actions, such as refining the query.
Previous research~\cite{JoosJKFPL24} shows that this browsing step remains a major bottleneck: it is time-consuming, error-prone, and places a high cognitive load on the user. Although the issues with browsing
keyframes are well-known, it has hardly been investigated by previous research. With our work, we contribute to more efficient browsing interfaces, outlined below.

Specifically, we investigate a retrieval scenario where users have a clear visual imagination of a target item. For example, they might remember a few seconds or a single frame of a video, and want to find its source. This scenario is widely discussed in the literature and academic competitions~\cite{JoosJKFPL24,vbs2023,vbs2024_eval_performance}, and is mimicked by the Visual Known-Item Search (KIS) task. For the Visual KIS task, a \emph{Known-Item}, such as a short video clip or a single frame, is presented to the participants - they then need to search for this target within a collection of videos. To simulate the browsing phase of the KIS task, we designed a large-scale study with 1715 tasks distributed among 49 participants, who were confronted with a specific target keyframe that they needed to locate in precomputed sets of candidate keyframes. By alternating between seven layout strategies, visualized in \autoref{fig:layout_comparison}, we conclude which layout is most efficient and effective for browsing.

The seven layouts were selected as distinct representatives of broad related work summarized in \autoref{sec:relatedWork}. 
The selection process involved various aspects such as human search behavior~\cite{organizedImages, JoosJKFPL24, stimulusSimilarity}, advanced arrangement options~\cite{Barthel2018,BarthelHKS23}, or effectiveness at competitions~\cite{vbs2024_eval_performance}. 
Motivated by applied systems in academic competitions~\cite{niiuit2,visione2024_novice,visione2024,lifeExplore2024,vibro2023}, we restrict ourselves to equal-sized keyframes on a fixed 2D grid.
More detailed discussion on motivations and selected layouts is presented in \autoref{sec:display_types}.
To challenge the layouts, we chose the highly homogeneous underwater Marine Video Kit (MVK)~\cite{MVK} dataset.
Despite its smaller size (28 hours resulting in 84309 keyframes), the homogeneity of the data and more abstract concepts (such as corals or the sea floor) make it challenging for users to formulate accurate queries \cite{vbs2024_eval_performance}.
Consequently, homogeneous datasets are characterized by long browsing phases and thus allow for comprehensive analysis of various browsing issues and patterns - for example, overlooks of the searched targets, scrolling behavior, or false skips of candidate lists containing the target image. 
Hence, in addition to a direct performance comparison of layouts, our analysis also provides novel insights into browsing phenomena related to the underlying data and layouts.

%We describe our data sampling procedure in Section XXX, alongside our study setup and participants.
%Our analysis of the collected data seeks to deepen the understanding of browsing effectiveness in video known-item search, providing insights into the following key questions.
% We could think of adding a contributions list here:
With our work, we investigate how the identification of target images in image grids can be advanced, making the following contributions:

\begin{itemize}
    \item A large-scale user study with 49 participants and 1715 tasks comparing seven keyframe layout strategies for the Visual Known-Item Search task in a homogeneous dataset.
    \item An integrated analysis of efficiency, effectiveness, and browsing behavior, including region skipping, scrolling, overlooks, and false skips.
    \item A thorough discussion on the implications of our findings for the design of retrieval systems, and guidance for future research opportunities, such as hybrid layouts and rank-preserving sorting.
    \item Source code available on GitHub\footnote{\url{https://github.com/Basti1211/vizkis-for-video-retrieval/tree/main}} that facilitates future research, and a deployed study environment\footnote{\url{https://viskis.org}} for reproducibility.
\end{itemize}

%% file: Chapter/2-related-work.tex
Representing videos effectively for browsing and retrieval is a long-lasting challenge.
Scanning through whole videos with a video player is time-consuming, and thus, most often only used to complement keyframe-based layouts~\cite{videoInterfacesReview, VideoSummaryLayout,keyframeIntroduction}.
These keyframes condense a video to the most important frames.
Approaches to visualizing keyframes on a screen are diverse and often aim to visualize only a single video a time.
Common techniques include linear side-by-side compositions~\cite{linearKeyframes}, video comics~\cite{KeyFrameComic}, hierarchical approaches~\cite{hierarchicalKeyframes, videoTree}, three-dimensional approaches~\cite{helixKeyframe}, and many more~\cite{videoInterfacesReview}.
Despite the diversity of the approaches, extensive studies comparing the effectiveness of layouts for video retrieval are missing. An exception is Joos et al.~\cite{JoosJKFPL24}, who compared two ranked-based layouts directly for video retrieval. Using eye-tracking, they compare the performance of 4-column layouts (where four images are displayed per row on a grid) to 8-column layouts (with eight images per row), and find that 4-column layouts are faster and more efficient, as they suffer from fewer overlooked target items. 

Due to the lack of extensive studies on video retrieval layouts, we review the related field of layouts for image retrieval. Compared to video retrieval, image retrieval provides a richer source of related literature that similarly focuses on the layout of images on the screen, but neglects possible video affiliation metadata.
While some approaches abstain from fixed-sized grids~\cite{ringImages, videoBrowsingInteractive,organizedImages, galaxyView}, there is also a substantial effort to optimize layouts on grid structures.
Conventional grid-based layouts ordered by rank are established, but research organizing candidates on a screen based on image similarity shows promising results~\cite{DynamicMaps, webClustering,BarthelHM15a}, and is theoretically founded on the Attentional Engagement theory by Duncan and Humphreys~\cite{stimulusSimilarity}. They find that search efficiency depends on the joint structure of target-distractor (T-D) and distractor-distractor (D-D) similarity, being most efficient when T-D is low and D-D is high, which enables the rapid rejection of homogeneous, non-matching regions. Low-level image features, such as color and texture~\cite{DynamicMaps, webClustering}, or high-level features, such as features extracted from neural networks~\cite{BarthelHM15a}, are utilized for image similarity measures.
These similarity measures are then incorporated in Graph-based algorithms and self-organizing maps for sorting, maximizing the similarity of neighboring images and minimizing the similarity of distant images~\cite{ImageGraph, BarthelHM15a}.
Initial results found that the organization of candidates facilitates the quick identification of relevant sub-regions, but might also lead to more overlooks~\cite{organizedImages}. 
%Our work contributes to this line of research by including layout functions that incorporate visual and semantic similarity.

Another body of research explicitly focuses on human search behavior in image and video browsing contexts with eye-tracking devices.
Xie et al. and Lu et al.~\cite{ExaminationBehavior,EyeTrackingImages} evaluate grid-based web image search and find a middle bias, indicating that users tend to focus more on the horizontal center than on the sides.
Concluding, conventional rank-based approaches that order images/videos from left to right and top to bottom place the best-matching candidate in columns that are more prone to being overlooked.
Furthermore, they find that users skip entire blocks of uninteresting regions. Joos et al.~\cite{JoosJKFPL24} confirm both the middle-bias and the row-skipping in an eye tracking study specifically designed for video retrieval. They further find that participants mostly skip those regions that have a high internal color similarity, but a low similarity to the target, confirming the Attentional Engagement theory. The reviewed work further agrees that participants follow a top-to-bottom search strategy. 
%We build upon these findings by including ranked layouts that explicitly utilize the middle bias, placing promising candidates into the center region.

Finally, we examine layout strategies of systems designed for competitive video retrieval. Therefore, we review tools from the Video Browser Showdown (VBS)~\cite{vbs2023, vbsEval} and the Lifelog Search Challenge (LSC)~\cite{lsc2024}. In these academic competitions, participants employ their tools to query video databases interactively, solving tasks such as the visual KIS task. While those competitions provide a holistic evaluation of video search tools, including efficient querying and browsing, our study focuses on isolating the performance of layouts in the browsing phase. In the context of these competitions, we found only one study that isolates the performance of keyframe layouts~\cite{NguyenNFH19}. Thereby, they compare a $k$-means-based layout to self-organizing maps, showing the superior performance of the latter. However, their study is only preliminary due to the low number of participants used. We fill this gap by providing a comprehensive study that includes multiple layouts of winning systems from the last few years.
NII-UIT~\cite{niiuit2}, the winner of VBS 2025, relies not only on keyframe ranks but also groups them into blocks of videos, similar to VISIONE, the winner of VBS 2024~\cite{visione2024_novice,visione2024}.
In contrast to that, lifeXplore~\cite{lifeExplore2024} won LSC 2024 with a ranked list of keyframes presented from top to bottom and left to right.
Yet another layout strategy was employed by VIBRO~\cite{vibro2023}, which won VBS 2023.
They leverage an efficient 2D grid layout algorithm~\cite{flas} to visualize keyframes sorted by their color. 
Further browsing strategies of VBS are summarized by Vadicamo et al.~\cite{vbs2024_eval_performance}. 

Despite the wide range of video retrieval systems, they largely focus on the process of retrieving a set of candidate images rather than their placement. A comprehensive evaluation comparing layouts based on search efficiency and effectiveness is still missing.

%% file: Chapter/3-study-setup.tex
\begin{figure}
    \centering
    \includegraphics[width=\linewidth]{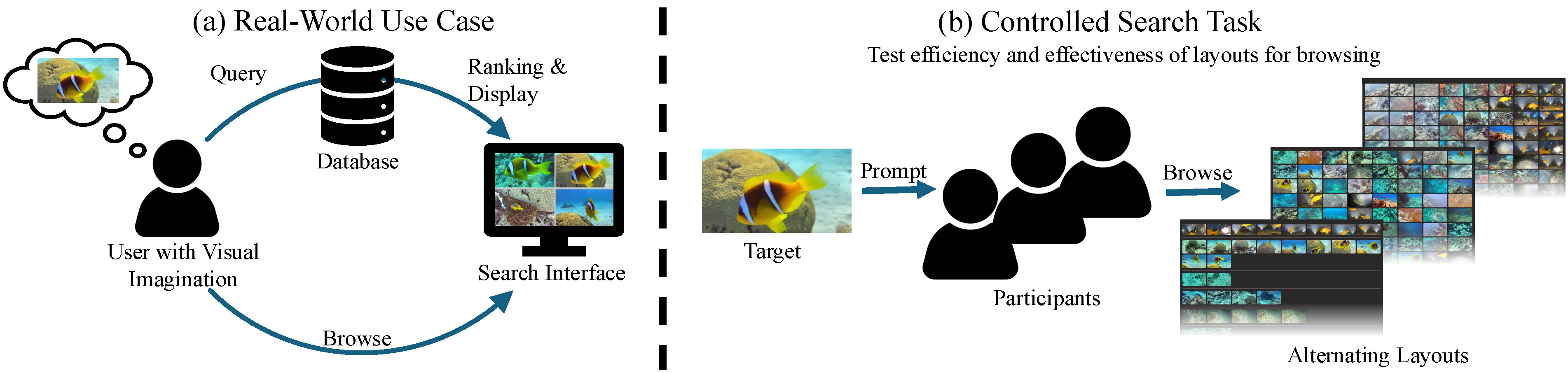}
    \caption{Motivation and study overview. Left: In real use, people try to find a specific video from memory and rely on iterative, visually driven browsing from partial visual cues. Right: We model this need as a controlled known-item search task: participants receive a target prompt and browse a large collection of keyframes ($|C_r|=200$) rendered with seven keyframe layouts, so we can isolate how layout design supports fast and precise re-identification during browsing.}
    \label{fig:use_case}
    \Description{This figure highlights how our study setup compares to real-world search scenarios, where users want to find a video source based on visual memory. The left panel sketches a real-world search pipeline. A user recalls a visual target (example: a yellow fish) and sends a query to a database. The system performs ranking and display, returning a screen of candidate keyframes. The user then browses the candidates, optionally refining the query in an iterative loop.
The right panel shows the controlled study setup that isolates the browsing component. Participants first see the target image explicitly, then search for that exact image among candidates displayed in different layouts. The same set of candidates is shown in multiple layout alternatives (for different participants) to measure how layout affects efficiency and effectiveness. The conceptual parallel between both panels is the visual inspection of a candidate pool given a known visual target.}
\end{figure}

As seen in the previous section, research on video retrieval systems has primarily focused on ranking the most relevant images in response to user queries.
Little attention has been given to the question of how the display of these images influences search efficiency and effectiveness, with rare exceptions~\cite{JoosJKFPL24}. 
Our work fills this gap for the Visual KIS task, where users need to locate a specific target (the \emph{Known-Item}) within a database. In the following, we formally introduce the display functions that generate keyframe layouts, motivate the selection of layouts, and outline the study setup.
\subsection{Compared Layout Strategies}
\label{sec:display_types}
Given a vector of ranked keyframe candidates $C_r$ returned from a search engine, display functions arrange $C_r$ on the screen. We thereby restrict ourselves to equal-sized keyframes and fixed 2D grids, motivated by the reviewed applied systems in VBS and LSC~\cite{niiuit2,visione2024_novice,visione2024,lifeExplore2024,vibro2023} and previous eye-tracking studies~\cite{JoosJKFPL24}. The first element $C_r[0]$ is assigned rank one and is expected to be the most relevant to the query. For a growing rank, the probability that an item is relevant gradually decreases. For our study, we assume that these ranks are given, and we fix the size of the candidate vector $C_r$ to 200 keyframes, as suggested by Joos et al.~\cite{JoosJKFPL24}. The display function can be formally defined as $f_d:[0, \dots, |C_r|) \rightarrow [0, \dots, cols-1] \times [0, \dots, rows-1]$, where $rows$ and $cols$ represent the number of rows and columns of the display grid. We reviewed in \autoref{sec:relatedWork} a rich body of proposed display functions. However, research on comparing which design decisions are favorable for efficiency and effectiveness for retrieval tasks is underexplored. An exception is Joos et al.~\cite{JoosJKFPL24}, who compared 4-column to 8-column layouts for the visual KIS task. However, their study was limited to only two ranked layouts, where the candidates' rank fully determines their final position. Our study continues this line of research and extends their setup in two ways: 1) We include the highly relevant class of grouped and sorted layouts, which are promising alternatives to ranked layouts and employed by winning systems of VBS and LSC. Specifically, grouped layouts, which cluster keyframes from the same video into contiguous regions to preserve context, were used by the winners of VBS2025 and VBS2024~\cite{niiuit2,visione2024_novice,visione2024}. Similarly, the winners of VBS2023 used sorted layouts, where keyframes are arranged by a chosen criterion, such as visual features, overriding strict rank order.~\cite{vibro2023} 2) We add two additional ranked layouts that incorporate findings of browsing behavior of Joos et al.~\cite{JoosJKFPL24}. Our layout pool lets us directly compare the rank-based configurations examined in prior work (and their refinements) with the newly added sorted and video-grouped variants.

We now formally introduce the seven layouts evaluated in our study: four rank-based layouts following Joos et al.~\cite{JoosJKFPL24}, plus two sorted and one video-grouped layout, both based on the VBS. We enrich the formal introductions with examples. For the examples, we assume that we only lay out $|C_r|=24$ keyframes - in our study, we used $|C_r|=200$.

\subsubsection{Ranked Layouts}
We present four types of ranked layouts, in which their rank fully determines the position of keyframes on the grid. For this section, we assume the ranks for each keyframe are already provided. We will describe how the candidate sets, including the ranks, were generated in \autoref{study_setup}.

\par\noindent
\begin{minipage}{\textwidth}
\begin{tcolorbox}[
  enhanced, breakable, sidebyside, sidebyside align=center,
  righthand width=0.4\textwidth,
  boxrule=0pt, frame hidden,
  colback=G4color!8!white,
  borderline west={2mm}{0pt}{G4color},
  left=6mm, right=2mm, top=2mm, bottom=2mm,
]
\textbf{Gradual 4 column grid (\GFour{})} -- an effective reference model according to the recent comparative study of Joos et al.~\cite{JoosJKFPL24}. The model uses a display function $f^{Gradual}_d(i) = [i \% 4, \lfloor i / 4 \rfloor]$ that organizes 200 ranked items into four columns and 50 rows. Gradual layouts are the most prominent display function in the VBS~\cite{vbs2023}. Although the display function is simple, Joos et al.~\cite{JoosJKFPL24} explored that the number of columns significantly influences perception: A four-column layout produced significantly fewer overlooks than an eight-column counterpart. Thus, although fewer keyframes were displayed on the screen, participants were faster with a four-column layout than with an eight-column layout.
\tcblower
\centering
\includegraphics[width=\linewidth]{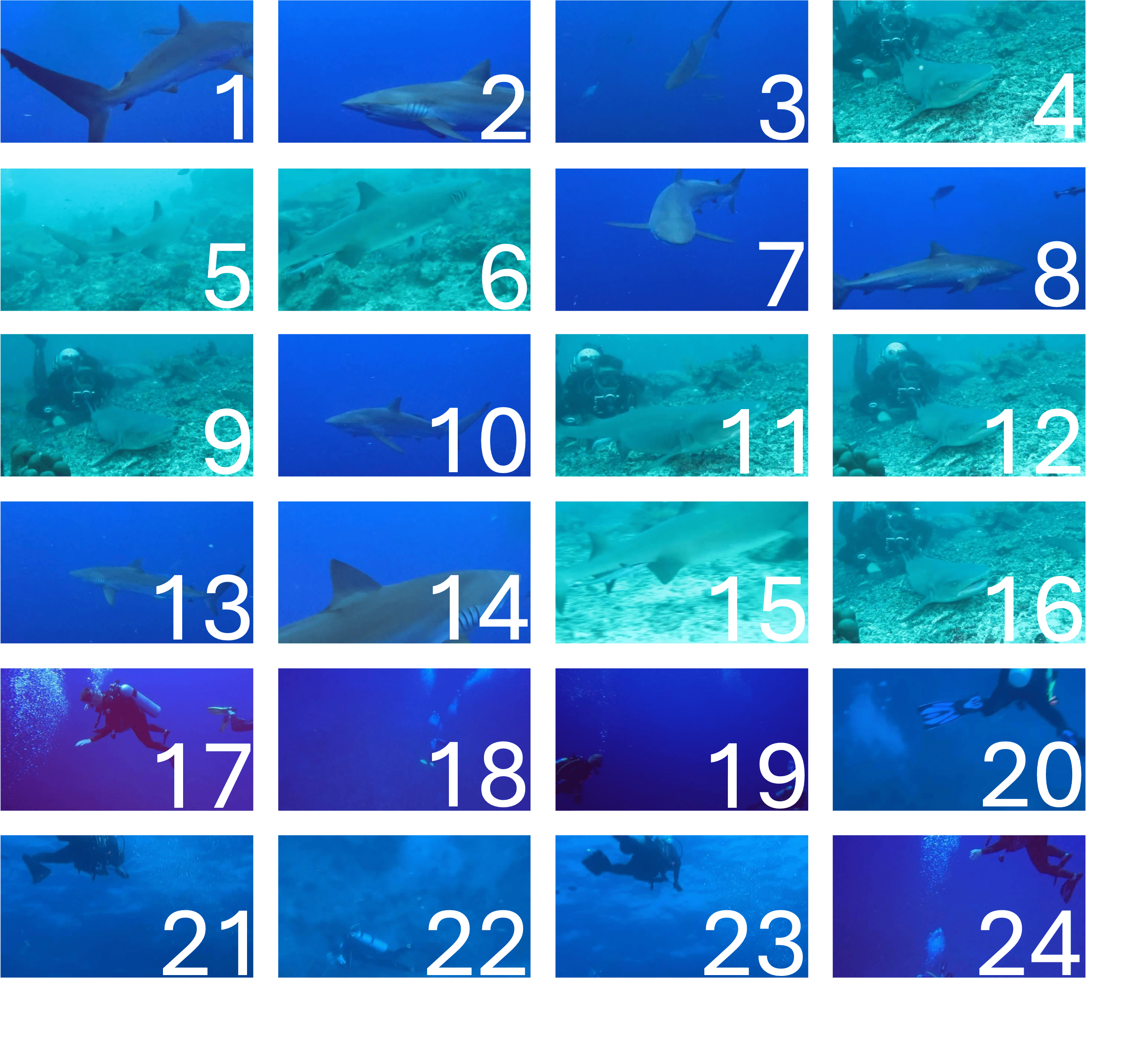}
\Description{This example shows 24 candidate thumbnails arranged in 4 columns × 6 rows. Each thumbnail is annotated with a rank number. Placement follows row-major order: ranks increase left-to-right within a row and then continue from the left of the next row. With only four columns, thumbnails are relatively large.}
\captionof{figure}{Example of \GFour{}. Keyframes are gradually displayed top-to-bottom, left-to-right in four columns.}
\end{tcolorbox}
\end{minipage}
\par

\par\noindent
\begin{minipage}{\textwidth}
\begin{tcolorbox}[
  enhanced, breakable, sidebyside, sidebyside align=center,
  righthand width=0.4\textwidth,
  boxrule=0pt, frame hidden,
  colback=G8color!8!white,
  borderline west={2mm}{0pt}{G8color},
  left=6mm, right=2mm, top=2mm, bottom=2mm
]
\textbf{Gradual 8 column grid (\GEight{})} -- a second reference model from~\cite{JoosJKFPL24} with a higher number of smaller images on one page, but lower number of rows to browse for a given candidate set. Specifically, the display function $f^{Gradual}_d(i) = [i \% 8, \lfloor i / 8 \rfloor]$ results in only 25 rows for 200 candidate items. 
\tcblower
\centering
\includegraphics[width=\linewidth]{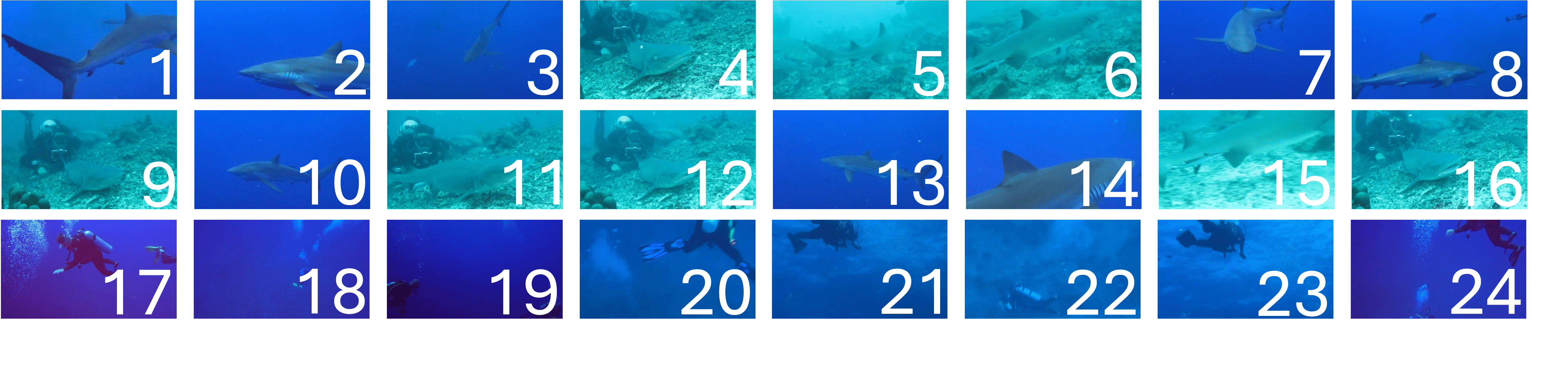}
\Description{This example shows the same 24 candidates as in Figure 3, arranged in 8 columns × 3 rows, following row-major order. Because the overall figure width is constant, thumbnails are smaller than in G4, trading per-item size for more items per row.}
\captionof{figure}{Example of \GEight{} using eight columns. Compared to \GFour{}, fewer rows are needed, but keyframes are smaller.}
\end{tcolorbox}
\end{minipage}

\par\noindent
\begin{minipage}{\textwidth}
\begin{tcolorbox}[
  enhanced, breakable, sidebyside, sidebyside align=center,
  righthand width=0.4\textwidth,
  boxrule=0pt, frame hidden,
  colback=C8color!8!white,
  borderline west={2mm}{0pt}{C8color},
  left=6mm, right=2mm, top=2mm, bottom=2mm
]
\textbf{Centric 8 column grid (\CEight{})} -- the recent eye-tracking study of Joos et al.~\cite{JoosJKFPL24} and others~\cite{ExaminationBehavior, EyeTrackingImages} suggested that the top left corner does not have to be an ideal place for the top-ranked item. Instead, the well-ranked items could be placed more towards the middle columns, as those profit from higher eye-gaze attention~\cite{JoosJKFPL24, ExaminationBehavior, EyeTrackingImages}. Instead, border columns are often overlooked and, consequently, should be filled with less relevant images - especially in the 8-column setting~\cite{JoosJKFPL24}. At the same time, 8-column layouts have the potential to display more images simultaneously. We build upon those two observations to construct an 8-column centric layout with a 2-4-2 pattern: The first two and last two columns are filled once the central four columns are filled. Specifically, $f^{Centric}_{d_1}(i) = [2 + i \% 4, \lfloor i / 4 \rfloor]$ is used for the first 100 elements, while the remaining 100 elements are assigned to the border columns.
\tcblower
\centering
\includegraphics[width=\linewidth]{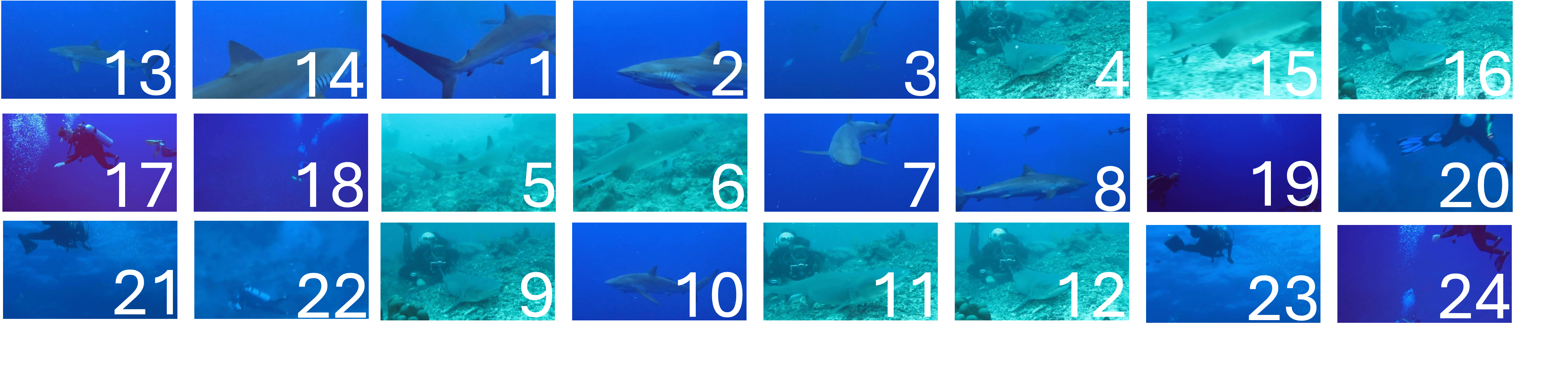}
\Description{This layout is also rank-based with eight columns, but it prioritizes the middle columns for higher ranks before filling towards the left and right edges. The intent is to align likely user attention with the spatial center (middle bias).}
\captionof{figure}{Example of \CEight{}. The most relevant keyframes are displayed in the four middle columns, where eye-gaze attention is highest~\cite{JoosJKFPL24, ExaminationBehavior, EyeTrackingImages}.}
\end{tcolorbox}
\end{minipage}
\par

\par\noindent
\begin{minipage}{\textwidth}
\begin{tcolorbox}[
  enhanced, breakable, sidebyside, sidebyside align=center,
  righthand width=0.4\textwidth,
  boxrule=0pt, frame hidden,
  colback=G4lpcolor!8!white,
  borderline west={2mm}{0pt}{G4lpcolor},
  left=6mm, right=2mm, top=2mm, bottom=2mm
]
\textbf{Gradual 4-column grid with left panel (\GFourLP{})} -- identical to \textbf{\GFour{}} but with a left-hand control panel that reduces the grid width and therefore thumbnail size. This reflects typical interactive search UIs with a persistent sidebar for navigation and filtering. Joos et al.~\cite{JoosJKFPL24} found that 4-column grids yield fewer overlooks than 8-column grids, likely due to fewer items per screen and larger thumbnails. \GFourLP{} preserves the 4-column structure while shrinking thumbnails, isolating the effect of keyframe size from column count. This lets us directly test whether thumbnail size alone impacts search performance. 
\tcblower
\centering
\includegraphics[width=\linewidth]{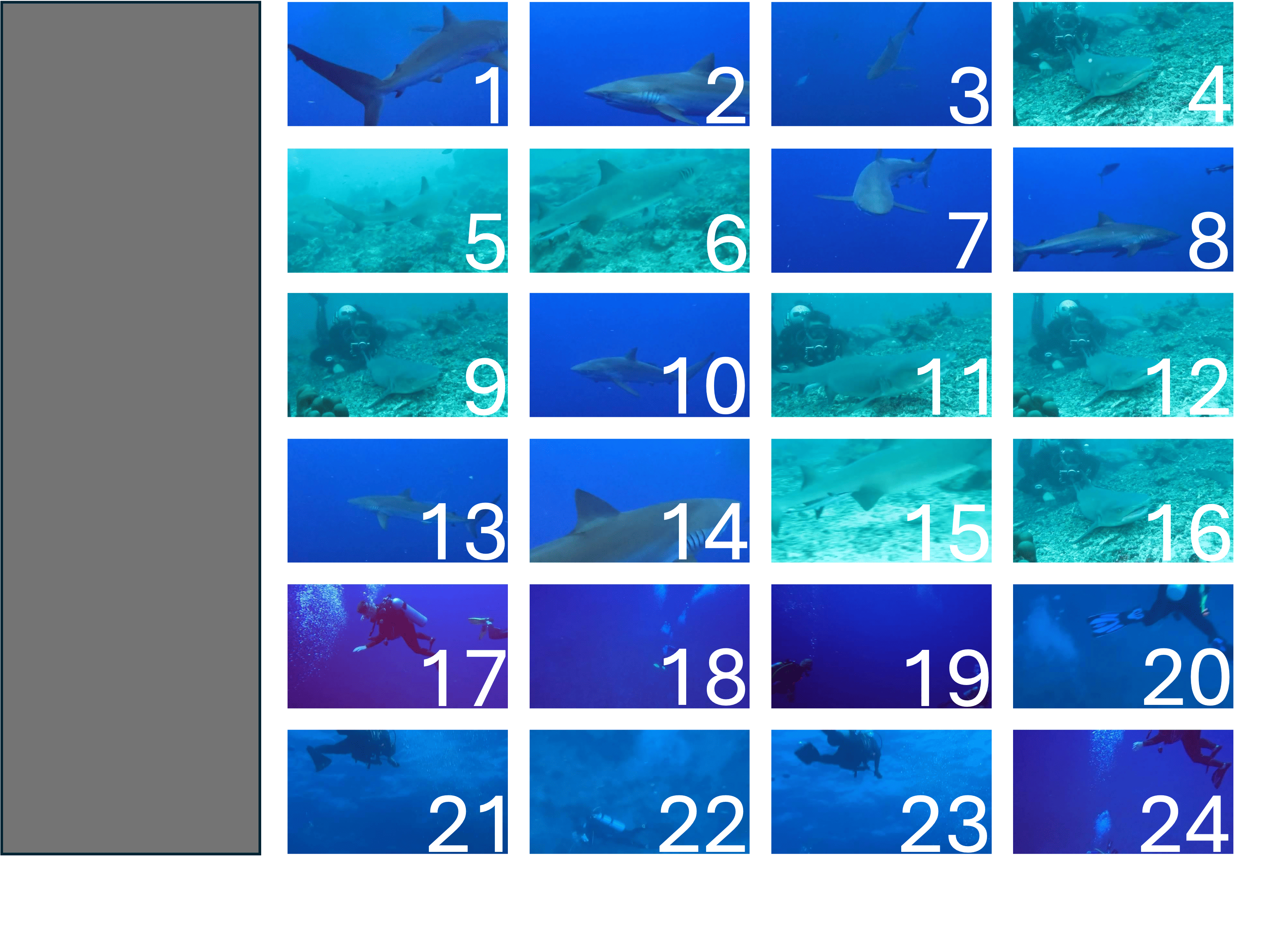}
\Description{This figure repeats the G4 structure (4 columns × 6 rows) but adds a side panel on the left for controls or metadata. Because the content area is narrower, the thumbnails are smaller than in Figure 3.}
\captionof{figure}{Example of \GFourLP{}. Due to a side panel, keyframes are displayed smaller.}
\end{tcolorbox}
\end{minipage}
\par 

\vspace{11px}

Summarizing, the four ranked layouts allow us to use \GFour{} and \GEight{} as reference models from Joos et al.~\cite{JoosJKFPL24}, while incorporating their key findings in two novel layouts: \GFourLP{} investigates the effect that bigger keyframes lead to fewer overlooks/better search performance, and \CEight{} tests whether the middle bias of eye-gaze attention can be used for increased search performance.

\subsubsection{Sorted Layouts}
In addition to the ranked layouts that map item rank directly to grid coordinates, we evaluate \emph{sorting layouts}. In these layouts, items are arranged so that neighboring grid cells contain images that are similar under an image–image similarity model~\cite{flas}; this strategy was also used by the VBS 2023 winners~\cite{vibro2023}. Joos et al.~\cite{JoosJKFPL24} observed that participants skip contiguous regions (in ranked layouts) whose internal similarity is high while similarity to the target is low. Sorting accentuates such homogeneous regions, potentially profiting from the skippable regions. Thereby, color similarity was better suited to explain these skipped regions than semantic similarity - we investigate which of the two is more effective for the overall search task. \\

\par\noindent
\begin{minipage}{\textwidth}
\begin{tcolorbox}[
  enhanced, breakable, sidebyside, sidebyside align=center,
  righthand width=0.4\textwidth,
  boxrule=0pt, frame hidden,
  colback=S8scolor!8!white,
  borderline west={2mm}{0pt}{S8scolor},
  left=6mm, right=2mm, top=2mm, bottom=2mm
]
\textbf{Sorted 8-column grid with semantic model (\TEightS{})} -- images are sorted in the grid so that semantically similar images are placed nearby in the grid. As the image-image similarity model, we use the cosine similarity with features from the CLIP model~\cite{SchallBHJ24} that was also utilized for initial text queries. Specifically, the sorting algorithm uses the FLAS approach\footnote{https://github.com/Visual-Computing/LAS\_FLAS} proposed by Barthel et al.~\cite{flas}. For reproducibility, we specify the hyperparameters in \autoref{sec:appendix_hyperparameters}.
\tcblower
\centering
\includegraphics[width=\linewidth]{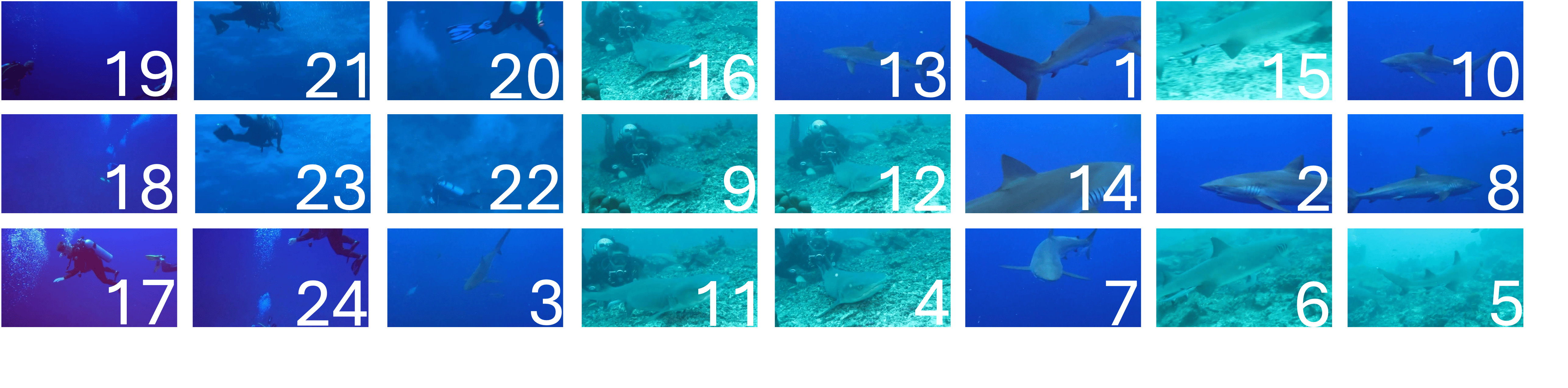}
\Description{The 24 exemplary candidates are displayed in 8 columns and sorted by semantic similarity rather than rank. As a result, conceptually related images appear next to each other. In the example, three loose clusters emerge from left to right: (1) divers, (2) divers with sharks, and (3) sharks without divers. Rank numbers no longer correspond to a strict order.}
\captionof{figure}{Example of \TEightS{}. Keyframes are sorted by their semantic content: On the left, keyframes of divers are visible, in the middle keyframes of divers and sharks are visible, and on the right, keyframes only with sharks are displayed.}
\end{tcolorbox}
\end{minipage}
\par

\par\noindent
\begin{minipage}{\textwidth}
\begin{tcolorbox}[
  enhanced, breakable, sidebyside, sidebyside align=center,
  righthand width=0.4\textwidth,
  boxrule=0pt, frame hidden,
  colback=S8ccolor!8!white,
  borderline west={2mm}{0pt}{S8ccolor},
  left=6mm, right=2mm, top=2mm, bottom=2mm
]
\textbf{Sorted 8-column grid with color-based model (\TEightC{})} -- we apply the same sorting algorithm as in \TEightS{}, but with LAB-color features. Specifically, we calculate for each image the mean of its color channels, leading to a 3-dimensional representation per image.
\tcblower
\centering
\includegraphics[width=\linewidth]{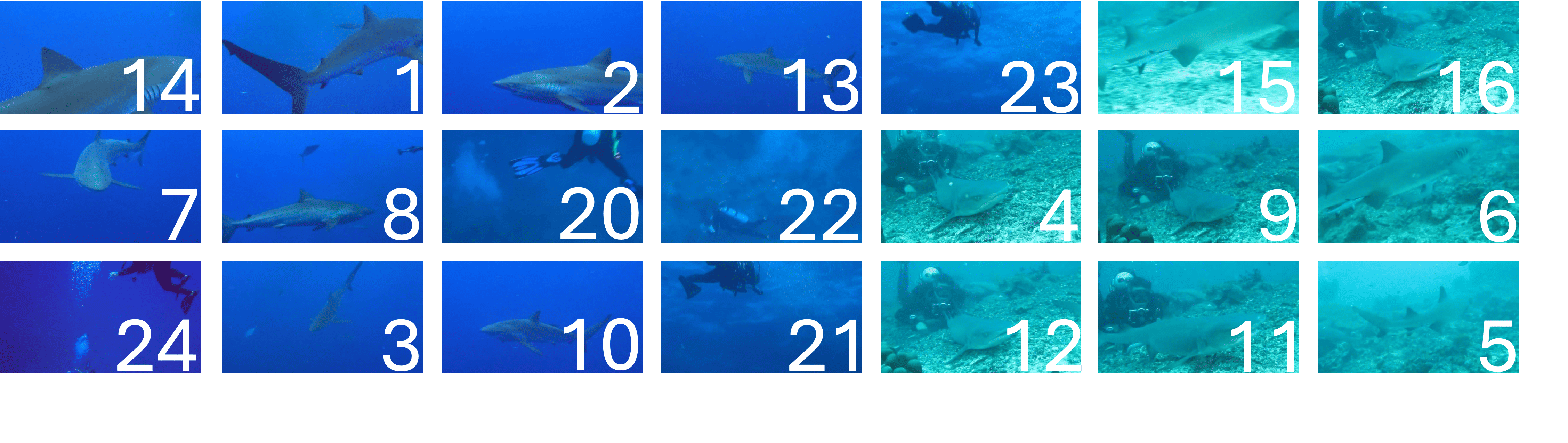}
\Description{Similar to Figure 7, but with color sorting. There are roughly two clusters visible: On the left, there are all darker blue images, while on the right, there are cyan images.}
\captionof{figure}{Example of \TEightC{}. Keyframes are sorted by color: on the left, all dark blue keyframes are displayed, while cyan keyframes are displayed on the right.}
\end{tcolorbox}
\end{minipage}
\par

\subsubsection{Grouped Layout} The last class of layout considers the video affiliation of the keyframes. Video grouping is the most common alternative to ranked layouts in VBS, for example, used by the winners of VBS2024~\cite{AmatoBCFGMVV24} and VBS2025~\cite{niiuit2}. While video grouping does not utilize the entire space of the 2D grid, it implicitly groups semantically and visually similar keyframes and showcases the temporal dimension of videos. \\
\par\noindent
\begin{minipage}{\textwidth}
\begin{tcolorbox}[
  enhanced, breakable, sidebyside, sidebyside align=center,
  righthand width=0.4\textwidth,
  boxrule=0pt, frame hidden,
  colback=V8color!8!white,
  borderline west={2mm}{0pt}{V8color},
  left=6mm, right=2mm, top=2mm, bottom=2mm
]
\textbf{Video grouping display (\VEight{})} -- items are grouped based on video ID, and the groups are sorted based on the relevance of the top-ranked item in the group. Items in each video group are sorted based on time. Similar to the skipping-regions behavior of sorted layouts, \VEight{} profits from the fact that whole groups of keyframes might be efficiently skipped, especially in homogeneous videos, where keyframes of one video are similar to each other.
\tcblower
\centering
\includegraphics[width=\linewidth]{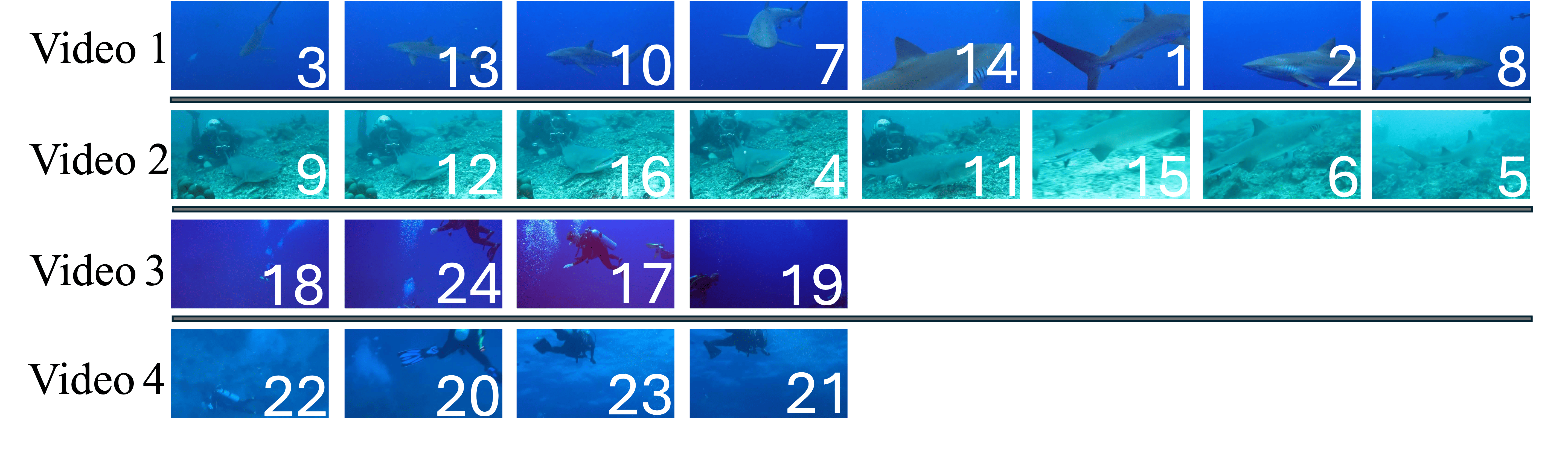}
\Description{Exemplary candidates are grouped by source video and arranged in blocks within an 8-column display. The example shows four videos containing 8, 8, 4, and 4 frames, respectively. Groups are ordered by the best-ranked keyframe contained in each group. Within a group, frames follow their temporal order. Because some groups do not fill complete rows, trailing cells in those rows remain blank.}
\captionof{figure}{Example of \VEight{}. Keyframes are displayed according to their video affiliation. In the example, keyframes originate from four videos.}
\end{tcolorbox}
\end{minipage}
\par
%and requires further extensions in future works. On the other hand, user studies have manpower and budget limitations, as well as significance correction issues. 
%Therefore, we did not sample a too large number of layout variants or their combinations. As it is difficult to guess (by intuition) the best possible display for a given task, and as the related work in this area is quite limited, we take the selected sample as a promising starting point opening doors to comprehensive analysis and conclusions that can be used when planning the next studies.

\subsection{Study Setup}\label{study_setup}

\subsubsection{Task and Implementation} We evaluate the performance of layouts for the Visual KIS task. In KIS, a target keyframe (or video, referred to as the \emph{Known-Item}) is provided, and participants must identify this target within a collection of videos. KIS simulates a scenario where users have a precise visual imagination of a scene that they want to find. In such situations, users describe their visual imagination to query a database, and are consequently confronted with candidates. Finally, they browse through these candidates to locate the target (see \autoref{fig:use_case}, left) or plan next steps to refine the query. 

In our study, we focus exclusively on the browsing phase. Each participant completes 35 browsing tasks that all follow the same structure. A target image (the \emph{Known-Item}) is presented to the participant in a web-based application. The participants must then locate this target in a collection of 200 candidate images (see \autoref{fig:use_case}, right). Each of the 35 tasks comprises different targets and candidates, described in the next subsection. After identifying a target among the candidates, it can be submitted as an answer. Correct answers lead to the next task, while incorrect answers prompt participants to retry. In contrast to Joos et al.~\cite{JoosJKFPL24}, we also prompted participants with tasks where the target is not in the candidates, a scenario that typically occurs in interactive video retrieval. Thus, participants can claim that the target is not in the collection, which will also transfer them to the next task. To ensure that participants do not skip prematurely, this option only becomes available after 30 seconds, which we identified as the minimum time required to search through the entire image collection. In our case, the target is a single keyframe instead of, for example, a short video clip. We decided on this setup for two reasons: first, we built on Joos et al.~\cite{JoosJKFPL24}, who also used keyframe prompts. Second, this setup better reflects real-world search behavior: people rarely maintain an entire video in working memory, so a single representative frame aligns with cognitive-load constraints.

Summarizing, one study run consists of 35 KIS tasks, where each of them is a unique combination of a (collection, layout) pair. By systematically alternating the layout across collections, we can evaluate layout-specific performances. We publish our application code and every generated (collection, layout) pair on GitHub\footnote{\url{https://anonymous.4open.science/r/viskis-for-video-retrieval-A50A}}. We furthermore deployed our ready-to-use study environment\footnote{\url{https://viskis.org}} for trial runs.

\subsubsection{Data}  To generate (target, candidate) collections, we chose the domain-specific marine video dataset MVK~\cite{MVK}, which has served as a benchmark in the VBS competition~\cite{vbs2023, vbsEval} and was also used by Joos et al.~\cite{JoosJKFPL24}. Both parts of the MVK collection, comprising a total of 28 hours of video resulting in 84309 keyframes, were utilized. We selected this homogeneous dataset because such collections typically yield weaker initial rankings, as visual concepts are harder to formulate in queries, resulting in longer browsing phases. This choice both reflects practical retrieval scenarios and poses a stronger challenge for layout design. To design the tasks, we extracted 35 collections, plus two additional collections for attention checks. We require that the collections are not too similar and simultaneously representative of the MVK dataset, leading to generalizable results for each layout. To this end, we followed the following six steps:

(1) In a preprocessing step, key representative frames were selected for every video in the MVK dataset, and CLIP-based embeddings~\cite{openclip, clip_2021} were generated for each of them. We use the same models and procedure as teams participating in the VBS~\cite{stroh2025prak}. (2) We run a clustering algorithm on the CLIP features of the keyframes. We use hierarchical clustering with Wards method ~\cite{ward1963}, not assuming specific shapes or densities of the underlying CLIP feature space. (3) We compute the centroid of 50 extracted clusters and get the closest representative images of it. (4) We annotate all 50 representative images with a natural language description and generate the text CLIP embeddings to simulate a real user query. (5) For each of the 50 text queries, we get the 200 best-matching keyframes. We filter near-duplicates in the time dimension, ensuring that every candidate in the collection is at least one second distant from each other. If we filter out the target, we set the candidate that caused the target to be filtered out as the new target. In cases where this happened, we manually examined whether the text query still matches the new target. (6) From the 50 collections, we randomly sample 35 (+2) collections. As some of the collections (11 out of 50) did not include the target based on the text description, we kept that balance in the final 35 collections, leading to 8 without a target.

\subsubsection{Study Design} For our study design, we had the following three requirements: First, every participant should be confronted with each layout and with each collection. Due to the large number of layouts and considerable variance among participants in previous studies~\cite{JoosJKFPL24}, a within-subject design is more feasible for our purpose. Second, every participant should have a balanced number of tasks per layout. As we investigated seven layouts, we distributed five tasks per layout to every participant, leading to 35 tasks, comparable to the number of tasks used in related work~\cite{JoosJKFPL24}. Our third requirement is that every task consisting of a (collection, layout) pair should, aggregated over participants, appear at the same frequency. To balance layout–collection pairs, we used a Latin-square schedule. We built a $7\times 35$ assignment matrix by concatenating five independently randomized $7\times 7$ Latin squares (one per block of seven collections). Rows are participants in a group of seven; columns are collections. Each block places every layout exactly once in each row and column, so across participants, each (collection, layout) pair occurs with equal frequency. After the assignment, we randomized the within-participant order of the 35 collections.

\subsubsection{Participants and Procedure} We recruited 49 participants (34 male, 14 female, 1 prefer not to tell), all with normal or corrected-to-normal vision (44.9\% of participants wore glasses or contact lenses). Their ages ranged from 19 to 47 years ($\mu = 25.6, \sigma=7.45$). No prior experience was required, and most participants had little to no background in video retrieval ($\mu = 1.7, \sigma=0.82$ on a five-point Likert scale, where '1' related to 'No Experience' and '5' to 'Expert-Level). 21 participants were recruited at a German university and 28 at a Czech university. Sessions in Germany used a 27-inch Dell UltraSharp U2722DE display; sessions in the Czech Republic used a 24-inch AOC e2460Sh. All tasks were solved using mouse and keyboard interaction. We discuss the effect of screen size in \autoref{appendix_screen_size}, and do not discriminate between screen sizes in the main paper.

Each participant followed the same procedure: After providing informed consent, they were given an overview of the task and the testing application via an introductory video. A brief question-and-answer session followed before participants started the tasks. We furthermore added attention checks after the 12th and 24th tasks - they consist of collections with only 20 candidate images that are easily solvable. Therefore, we could measure whether participants skipped tasks without trying to solve them. At the end, participants filled out a questionnaire about demographics and prior knowledge, and received a compensation of 10 EUR or 200 CZK, respectively. 

\subsubsection{Relation of the task to real search scenario}
In real visual known-item search scenarios, it is expected that users can recognize the searched target keyframe. However, in the laboratory evaluation setting where users solve 35 tasks in one hour, it is difficult to memorize all target keyframes and recognize them in the result set. Hence, inspired by VBS/LSC competitions~\cite{vbsEval,vbs2024_eval_performance,lsc2024}, we allowed participants to look at the target image on demand anytime during the search. In the testing application, users can open a window with the target image while the background is blurred. This oracle feature was the same for all display types tested. The increase in search time caused by these views is related to the need to recall the target keyframe, which is also affected by the type of display used. Another ``oracle'' feature was implemented to the detail view of a selected keyframe, where users can view a larger version of images. Next to the larger thumbnail of the selected keyframe, we also show the target image, so that the users can identify the target more easily. By showing the target image, we tried to fill the memorization gap caused by a large number of evaluated tasks, as well as highly visually similar content in the dataset. 

%Therefore, all measured search times should be considered as approximates affected by the study design. Besides that, it is interesting to compare the numbers of requested target views and detail views for different types of displays.

%% file: Chapter/4-results.tex
Each of our 49 participants solved 35 tasks, resulting in 1715 solved (layout, collection) pairs.
Consequently, each layout was tested 245 times in total and seven times for each collection. In the following, we share findings generalizable to the layout and discuss insights related to underlying image collections.

\autoref{tab:overview} offers an overview of the most critical metrics measured. We additionally split our evaluation into two groups, where collections either contained the target or did not. We observe that this heavily influences which of the layouts is favorable.

\begin{table*}
\caption{The left part contains comparison of layouts with averaged times over all tasks for four measurements. $\uparrow$/$\downarrow$ indicate better/worse performance. \underline{Time} is the overall search time from task start to submit/skip, \underline{F. A.} is the first-arrival time (first time target appeared on the screen), \underline{T. V.} is the duration for how long the target overlay was activated, \underline{Comp.} describes how long the image detail and comparison overlay were visible. The three columns in the right part of the table shows the chance for a specific event to occur in a task (i.e., |events| / |tasks|). \underline{Wrongs} measures wrongly submitted candidates, \underline{F. S.} is the number of false skips, \underline{Overl.} is the number of overlooks of the target, \underline{Rel. Scroll} is the relative amount of scrolling with respect to the height of the grid, e.g., 100\% scrolling describes a participant scrolling once from top to bottom.}
\centering
\small
\renewcommand{\arraystretch}{1.2}
\setlength{\tabcolsep}{5pt}

\begin{tabular}{lcccc|cccc}
\toprule
\textbf{Condition} &
\textbf{Time (s) $\downarrow$} &
\textbf{F. A. (s) $\downarrow$} &
\textbf{T. V. (s)} &
\textbf{Comp. (s)} &
\textbf{Wrongs (\%) $\downarrow$} &
\textbf{F. S. (\%) $\downarrow$} &
\textbf{Overl. (\%) $\downarrow$} &
\textbf{Rel. Scroll (\%)} \\
\midrule
\multicolumn{9}{l}{\textbf{Total}} \\
\GFour{}     & 42.7 & 8.1 & 2.8 & 6.0 & 4.5 & 3.7 & \textbf{20.1} & 80.1 \\
\GFourLP{}   & 42.3 & 6.9 & 3.0 & 5.9 & 4.1 & \textbf{1.6} & 21.2 & 79.3 \\
\GEight{}    & 42.5 & \textbf{3.5} & 3.1 & 6.6 & \textbf{3.3} & 4.2 & 27.5 & 78.7 \\
\CEight{}    & 44.0 & 6.4 & 3.0 & 7.1 & 4.5 & 5.8 & 33.3 & 87.1 \\
\TEightC{}   & 44.4 & 12.6 & 3.2 & 7.6 & 9.0 & 6.3 & 34.4 & 117.6 \\
\TEightS{}   & 42.7 & 9.8 & 3.0 & 6.8 & 6.5 & 4.8 & 40.7 & 102.8 \\
\VEight{}    & \textbf{40.1} & 4.4 & 3.2 & 6.4 & \textbf{3.3} & 3.7 & 35.4 & 88.0 \\
\midrule
\multicolumn{9}{l}{\textbf{Target in Collection}} \\
\GFour{}     & 25.1 & 8.1 & 2.3 & 6.0 & 5.8 & 3.7 & \textbf{20.1} & 39.9 \\
\GFourLP{}   & \textbf{24.3} & 6.9 & 2.3 & 5.4 & 4.8 & \textbf{1.6} & 21.2 & 41.7 \\
\GEight{}    & 25.0 & \textbf{3.5} & 2.7 & 5.6 & \textbf{3.2} & 4.2 & 27.5 & 35.3 \\
\CEight{}    & 29.0 & 6.4 & 2.7 & 6.7 & 4.2 & 5.8 & 33.3 & 47.2 \\
\TEightC{}   & 32.3 & 12.6 & 2.7 & 7.6 & 11.6 & 6.3 & 34.4 & 73.1 \\
\TEightS{}   & 32.0 & 9.8 & 2.7 & 6.8 & 7.9 & 4.8 & 40.7 & 68.9 \\
\VEight{}    & 25.1 & 4.4 & 2.8 & 5.6 & \textbf{3.2} & 3.7 & 35.4 & 46.4 \\
\midrule
\multicolumn{9}{l}{\textbf{Target not in Collection}} \\
\GFour{}     & 102.2 & - & 4.7 & 5.8 & \textbf{0.0} & - & - & 215.8 \\
\GFourLP{}   & 103.1 & - & 5.2 & 7.6 & 1.8 & - & - & 206.1 \\
\GEight{}    & 101.5 & - & 4.2 & 10.0 & 3.6 & - & - & 225.1 \\
\CEight{}    & 94.6  & - & 3.8 & 8.5 & 5.4 & - & - & 221.8 \\
\TEightC{}   & 85.1  & - & 5.0 & 7.4 & \textbf{0.0} & - & - & 267.8 \\
\TEightS{}   & \textbf{78.8} & - & 3.7 & 7.0 & 1.8 & - & - & 217.1 \\
\VEight{}    & 90.6  & - & 4.5 & 8.9 & 3.6 & - & - & 228.5 \\
\bottomrule
\end{tabular}
\label{tab:overview}
\end{table*}

\subsection{Efficiency}\label{sec:efficiency}
The efficiency of layouts is mainly measured by the overall search time, defined as the time it took a participant from starting a task (after inspecting the target) to finish it. Average values per layout are reported in \autoref{tab:overview}, whereas we report on statistical significance in  \autoref{tab:significance_search_time}. We applied pairwise Wilcoxon Signed-Rank tests, as preliminary Shapiro-Wilk tests indicated non-normal distributions. To control for the family-wise error rate resulting from multiple comparisons, we further applied the Holm-Bonferroni correction. They reveal insignificant differences in the overall search time between layouts but significant differences when differentiating whether collections contain a target. While the video grouped layout \VEight{} achieves the fastest overall performance, \GFourLP{} is the fastest to find a target in a collection, and \TEightS{} is the fastest to confirm that a target is not in the collection. Furthermore, sorted layouts perform generally worse than all ranked-based layouts for collections with a target, but outperform them if collections do not contain a target. We investigate user search behavior and dataset characteristics that lead to these findings below.

\begin{table*}
\caption{$p$-values for pairwise Wilcoxon Signed-Rank tests for the search time. The lower triangle represents the Holm-Bonferroni corrected $p$-values, while the upper triangle shows the uncorrected $p$-values.}
\centering
\renewcommand{\arraystretch}{1.2}
\setlength{\tabcolsep}{5pt}

\begin{tabular}{l|ccccccc}

\toprule
\multicolumn{8}{c}{\textbf{Total}}\\
\midrule
      & \GFour{} & \GFourLP{} & \GEight{} & \CEight{} & \TEightC{} & \TEightS{} & \VEight{} \\
\midrule
\GFour{}    & - & 0.716 & 0.602 & 0.720 & 0.165 & 0.302 & 0.723 \\
\GFourLP{}  & 1.000 & - & 0.708 & 0.382 & 0.096 & 0.328 & 0.962 \\
\GEight{}   & 1.000 & 1.000 & - & 0.411 & 0.059 & 0.130 & 0.937 \\
\CEight{}   & 1.000 & 1.000 & 1.000 & - & 0.684 & 0.899 & 0.433 \\
\TEightC{}  & 1.000 & 1.000 & 1.000 & 1.000 & - & 0.417 & 0.109 \\
\TEightS{}  & 1.000 & 1.000 & 1.000 & 1.000 & 1.000 & - & 0.296 \\
\VEight{}   & 1.000 & 1.000 & 1.000 & 1.000 & 1.000 & 1.000 & - \\
\midrule
\toprule
\multicolumn{8}{c}{\textbf{Target in Collection}}\\
\midrule
      & \GFour{} & \GFourLP{} & \GEight{} & \CEight{} & \TEightC{} & \TEightS{} & \VEight{} \\
\midrule
\GFour{}    & - & 0.789 & 0.686 & 0.211 & \textbf{<0.001} & \textbf{<0.001} & 0.247 \\
\GFourLP{}  & 1.000 & - & 0.745 & 0.092 & \textbf{<0.001} & \textbf{0.001} & 0.133 \\
\GEight{}   & 1.000 & 1.000 & - & 0.072 & \textbf{<0.001} & \textbf{<0.001} & 0.140 \\
\CEight{}   & 1.000 & 0.918 & 0.793 & - & \textbf{0.028} & 0.062 & 0.473 \\
\TEightC{}  & \textbf{0.006} & \textbf{<0.001} & \textbf{<0.001} & 0.361 & - & 0.824 & \textbf{0.001} \\
\TEightS{}  & \textbf{0.009} & \textbf{0.021} & \textbf{0.016} & 0.748 & 1.000 & - & \textbf{<0.001} \\
\VEight{}   & 1.000 & 1.000 & 1.000 & 1.000 & \textbf{0.016} & \textbf{0.016} & - \\
\midrule
\toprule
\multicolumn{8}{c}{\textbf{Target not in Collection}}\\
\midrule
      & \GFour{} & \GFourLP{} & \GEight{} & \CEight{} & \TEightC{} & \TEightS{} & \VEight{} \\
\midrule
\GFour{}    & - & 0.726 & 0.237 & 0.264 & \textbf{0.012} & \textbf{0.002} & \textbf{0.047} \\
\GFourLP{}  & 1.000 & - & 0.671 & 0.396 & \textbf{0.011} & \textbf{0.002} & \textbf{0.041} \\
\GEight{}   & 1.000 & 1.000 & - & 0.666 & 0.108 & \textbf{0.012} & 0.257 \\
\CEight{}   & 1.000 & 1.000 & 1.000 & - & 0.112 & \textbf{0.011} & 0.237 \\
\TEightC{}  & 0.208 & 0.208 & 1.000 & 1.000 & - & 0.285 & 0.181 \\
\TEightS{}  & \textbf{0.032} & \textbf{0.032} & 0.208 & 0.208 & 1.000 & - & 0.125 \\
\VEight{}   & 0.664 & 0.609 & 1.000 & 1.000 & 1.000 & 1.000 & - \\
\bottomrule
\end{tabular}
\label{tab:significance_search_time}
\end{table*}

\subsubsection{Position loss of sorted layouts} Sorted approaches do not preserve the initial ranking based on text queries. As most target ranks are low, they are positioned towards the first (upper) rows of the grid in ranked-based layouts. 63\% of targets have a lower rank than 20, and 77.7\% of targets have a lower rank than 50. We show an exact distribution in \autoref{supplementary_rank_pos}. We remind that the distribution is not artificial and is the result of text queries describing target images. In the sorted approaches, the low-ranked targets lose their favorable position toward the top of the grid, leading to an average row loss of 10.6 for \TEightS{} and 9.2 for \TEightC{}, compared to \GEight{}. Thus, targets are positioned roughly one-third further down in the grid. Consequently, those targets are observed later as participants start to scan the grid from top to bottom. We provide evidence for that by measuring the time when the target appears first on the screen (``First Arrival'' in \autoref{tab:overview}). For the eight-column layouts, the magnitude of differences in the first-arrival times is similar to the differences in the overall search time. Furthermore, the rank loss leads to a higher relative scroll rate for sorted layouts.
\autoref{tab:rank_search_time} provides further evidence for the effect of target ranks on search efficiency by binning collections based on their rank. Specifically, we report average search times separately for target rank/position below or above 56 (after row 7/14 for eight/four-column layouts where the initial page without scrolling ends for 8-column layouts). Thereby, the position on the grid is defined in row-major order, read left to right and top to bottom. The overall differences in search times result primarily from a preferable target position of ranked layouts:  \TEightS{}, \TEightC{}, and \VEight{} are considerably faster than the ranked-based alternatives when comparing them based on the position in the grid. We emphasize that this is a structural disadvantage of sorted grids and not of our experimental setting: For distributions where the target is likely to have a low rank, sorting layouts might push targets to a less-favorable position.

\begin{table}[ht]
\caption{Average Task completion time based on the target rank and position.}
\centering

\begin{tabular}{c c c c c c c c c}
\hline
\textbf{Layout} & \textbf{\GFour{}} & \textbf{\GFourLP{}} & \textbf{\GEight{}} & \textbf{\CEight{}} & \textbf{\TEightC{}} & \textbf{\TEightS{}}  & \textbf{\VEight{}}\\
\hline
Rank $\leq$ 56 & 20.4 & 20.5 & 20.5 & 24.5 & 30.1 & 31.7 & 22.5 \\
Rank $>$ 56 & 41.7 & 37.8 & 41.0 & 44.7 & 40.3 & 32.7 & 34.0\\
Pos.  $\leq$ 56 & 20.4 & 20.5 & 20.5 & 21.6 & 11.0 & 16.7 & 22.9\\

Pos.  $>$ 56 & 41.7 & 37.8 & 41.0 & 43.8 & 34.1 & 35.4 & 28.8 \\
\hline
\end{tabular}
\label{tab:rank_search_time}
\end{table}

\subsubsection{Overlooks} The second disadvantage of sorted layouts is the higher number of overlooks. While 4-column layouts also suffer from higher first-arrival times (as fewer images are displayed on the screen), they counter this effect by fewer overlooks, in line with Joos et al.~\cite{JoosJKFPL24}. We define overlooks as cases where a target is fully displayed on the viewport, and subsequently leaves it. The time passing between an overlook's first appearance and task completion is 45.1 seconds on average, almost twice as long as the average overall task completion time. Based on pairwise Wilcoxon Signed-Rank tests with Holm-Bonferroni corrections, the number of overlooks between methods are not significant (see \autoref{supplementary_overlooks} for exact $p$-values). However, due to the long delays in search times, we assess their impact on the overall search time as substantial. Thereby, not only the size of displayed keyframes is crucial for preventing overlooks: Although keyframes in \GFourLP{} are displayed substantially smaller than in \GFour{}, overlooks appear similarly often. We infer that not only the size of keyframes leads to more overlooks, but also the number of (similar) keyframes displayed on a screen.

\begin{figure}[t]
    \centering
\includegraphics[width=1\linewidth]{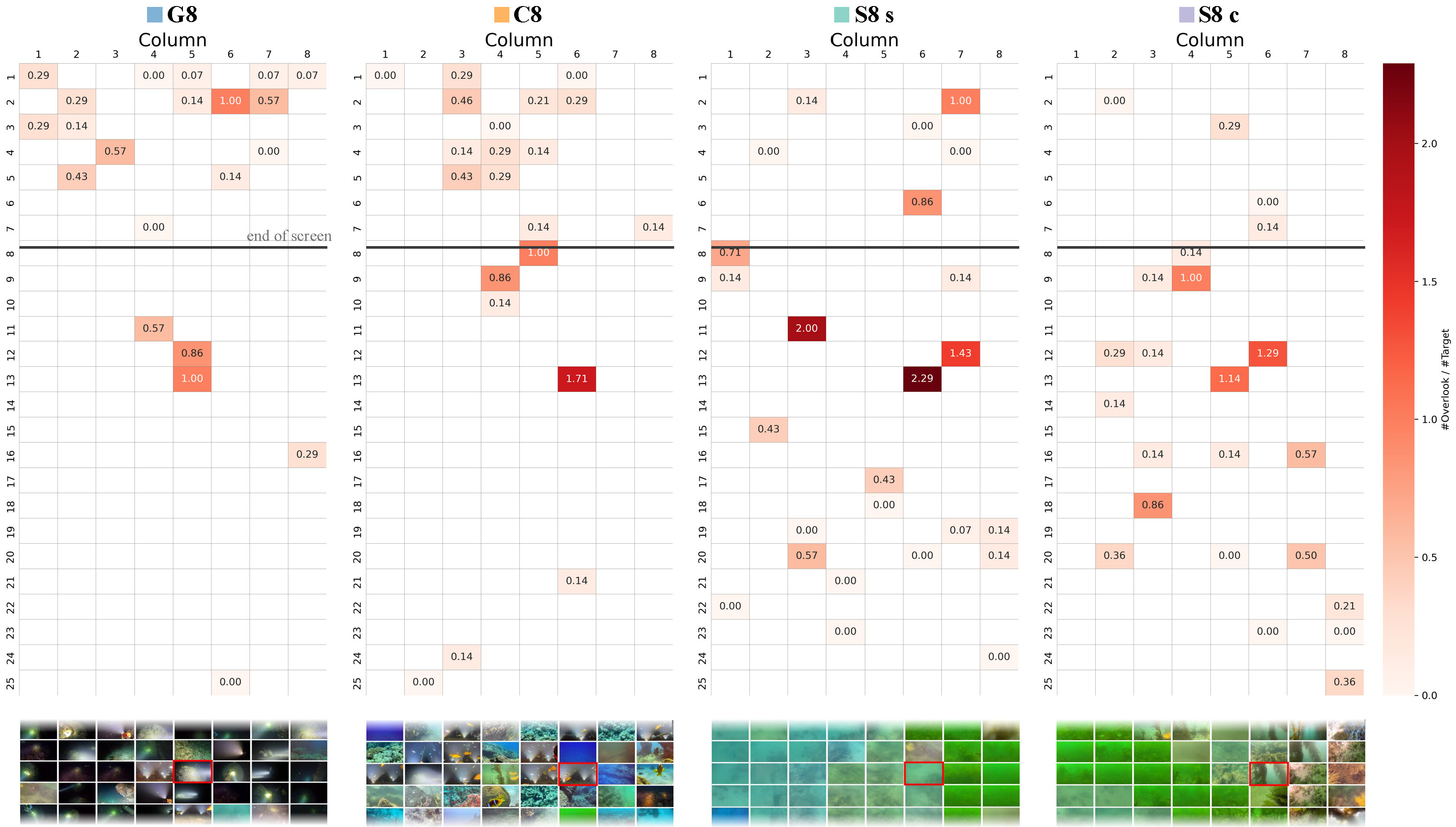}
    \caption{Heatmap of average overlooks per grid cell by target position. The bold line marks the display boundary. We also show for each layout the section with the most overlooked target (marked red).}
    \Description{This figure compares all 8-column layouts except video grouping using heatmaps of overlooks for each cell in the grid. Overlooks concentrate in the vertical middle of the grid and are less common at the top and bottom, suggesting that scrolling may contribute to misses. In contrast to prior middle-bias expectations, the figure does not show a consistent horizontal-center effect across columns. We furthermore plot the most overlooked targets and their neighborhood for each layout, respectively. The neighborhoods show visually similar candidates to the targets.}
    \label{fig:heatmap_overlooks}
\end{figure}

\newcommand{\rowlabel}[1]{\multicolumn{10}{@{}l}{\textbf{#1}}\\}

\begin{table}[h]
\caption{Probability that target is overlooked and task was solved given that target is in the column. \emph{Overall} aggregates all 8-column layouts except \VEight{}. Best values are marked \textbf{bold}, while worst values are marked \textcolor{blue}{blue}. "-" indicates no targets are in the column.}
\centering
\begin{tabular}{@{}l@{\,$\mid$\,}l*{8}{c}@{}}
\toprule
\multicolumn{2}{l}{\textbf{Column $i$}} & 1 & 2 & 3 & 4 & 5 & 6 & 7 & 8 \\
\midrule

\rowlabel{Overall}
$P(Overlook$ & $Target\;in\;Column\;i)\downarrow$  & 0.2449 & 0.2208 & 0.4286 & 0.2738 & 0.3383 & \textcolor{blue}{0.5089} & 0.3846 & \textbf{0.1667} \\
$P(Solved$ & $Target\;in\;Column\;i)\downarrow$     & 0.9592 & \textcolor{blue}{0.9351} & 0.9524 & 0.9405 & 0.9398 & 0.9375 & 0.9560 & \textbf{0.9643} \\
\midrule

\rowlabel{\GEight{}}
$P(Overlook$ & $Target\;in\;Column\;i)\downarrow$  & 0.2857 & 0.2857 & \textcolor{blue}{0.5714} & 0.1904 & 0.3265 & 0.3809 & 0.1785 & \textbf{0.1428} \\
$P(Solved$ & $Target\;in\;Column\;i)\downarrow$       & 0.9523 & 0.9523 & \textbf{1.0} & \textbf{1.0} & \textcolor{blue}{0.9183} & 0.9523 & 0.9642 & \textbf{1.0} \\
\midrule

\rowlabel{\CEight{}}
$P(Overlook$ & $Target\;in\;Column\;i)\downarrow$  & \textbf{0.0} & \textbf{0.0} & 0.3492 & 0.3143 & 0.3429 & \textcolor{blue}{0.4857} & - & 0.1429 \\
$P(Solved$ & $Target\;in\;Column\;i)\downarrow$      & \textbf{1.0} & \textbf{1.0} & 0.9524 & \textcolor{blue}{0.8857} & 0.9429 & 0.9429 & - & \textbf{1.0} \\
\midrule

\rowlabel{\TEightS{}}
$P(Overlook$ & $Target\;in\;Column\;i)\downarrow$  & 0.2857 & 0.2143 & 0.6786 & \textbf{0.0} & 0.2857 & \textcolor{blue}{0.7857} & 0.4524 & 0.0950 \\
$P(Solved$ & $Target\;in\;Column\;i)\downarrow$      & 0.9524 & \textbf{1.0} & 0.9286 & \textbf{1.0} & \textbf{1.0} & \textcolor{blue}{0.8929} & 0.9286 & \textbf{1.0} \\
\midrule

\rowlabel{\TEightC{}}
$P(Overlook$ & $Target\;in\;Column\;i)\downarrow$ & - & \textbf{0.2286} & 0.3214 & \textcolor{blue}{0.5714} & 0.3929 & 0.3571 & 0.5238 & \textbf{0.2286} \\
$P(Solved$ & $Target\;in\;Column\;i)\downarrow$        & - & \textcolor{blue}{0.8857} & 0.9643 & 0.9286 & 0.9286 & 0.9643 & \textbf{1.0} & 0.9143 \\
\bottomrule
\end{tabular}
\label{tab:overlooks_per_column}
\footnotetext{“–” indicates no targets occurred in that column; value not estimable.}
\end{table}

Additionally, we investigated where overlooks are most likely to appear. Previous eye tracking studies have shown a middle bias; in their studies, participants' gazes focused more on the center columns, which may explain the overlooks of targets in outer columns. \autoref{tab:overlooks_per_column} does not confirm this middle bias. Contrarily, we find a scattered distribution without a clear trend. While the outer columns are, overall, less overlooked than the middle columns, the distribution of probabilities varies highly between layouts. The probability that a task was successfully solved by finding the target also does not confirm a clear middle bias. Our hypothesis is also supported by the bad performance of \CEight{}: Although more keyframes are placed towards the middle columns compared to \GEight{}, search time and correctness are worse, and overlooks still appear similarly often.

\autoref{fig:heatmap_overlooks}, where we plot the overlooks per grid cell, reveals two more findings: Firstly, overlooks do not appear more often in the \emph{horizontal} center, but in the \emph{vertical} center. As targets at the bottom of the grid are less overlooked, we hypothesize that scrolling negatively impacts overlooks. Fujii et al.~\cite{blurryScrcolling2} outline that scrolling might impair visual perception, either because eyes are failing to keep up with the moving target, or because information from images is mentally overwritten with new information. Secondly, we find a strong aggregation of overlooks for single tasks. We hypothesize that the visual characteristics of the target and its neighborhood influence the overlooks more than its position on the grid. We plot the most overlooked sections for each layout below their heatmap, respectively. While future studies are necessary to explore the influence of visual characteristics on overlooks in more detail, we find that similar candidates close to the target might have influenced participants. A possible explanation for this could be the effect of visual crowding~\cite{visCrowding}, where similar-looking entities impair peripheral vision - participants could more often overlook the correct target as their peripheral system pointed them to visually similar close candidates instead of the correct target. This type of crowd filtering error might also be a potential disadvantage for grid-sorting algorithms. The objective of grid functions favors locally homogeneous clusters, leading to targets that might be placed alone near a cluster recognized as incorrect.

\subsubsection{Skipping Regions}
\begin{figure}[p]
    \centering
    \includegraphics[width=0.88\linewidth]{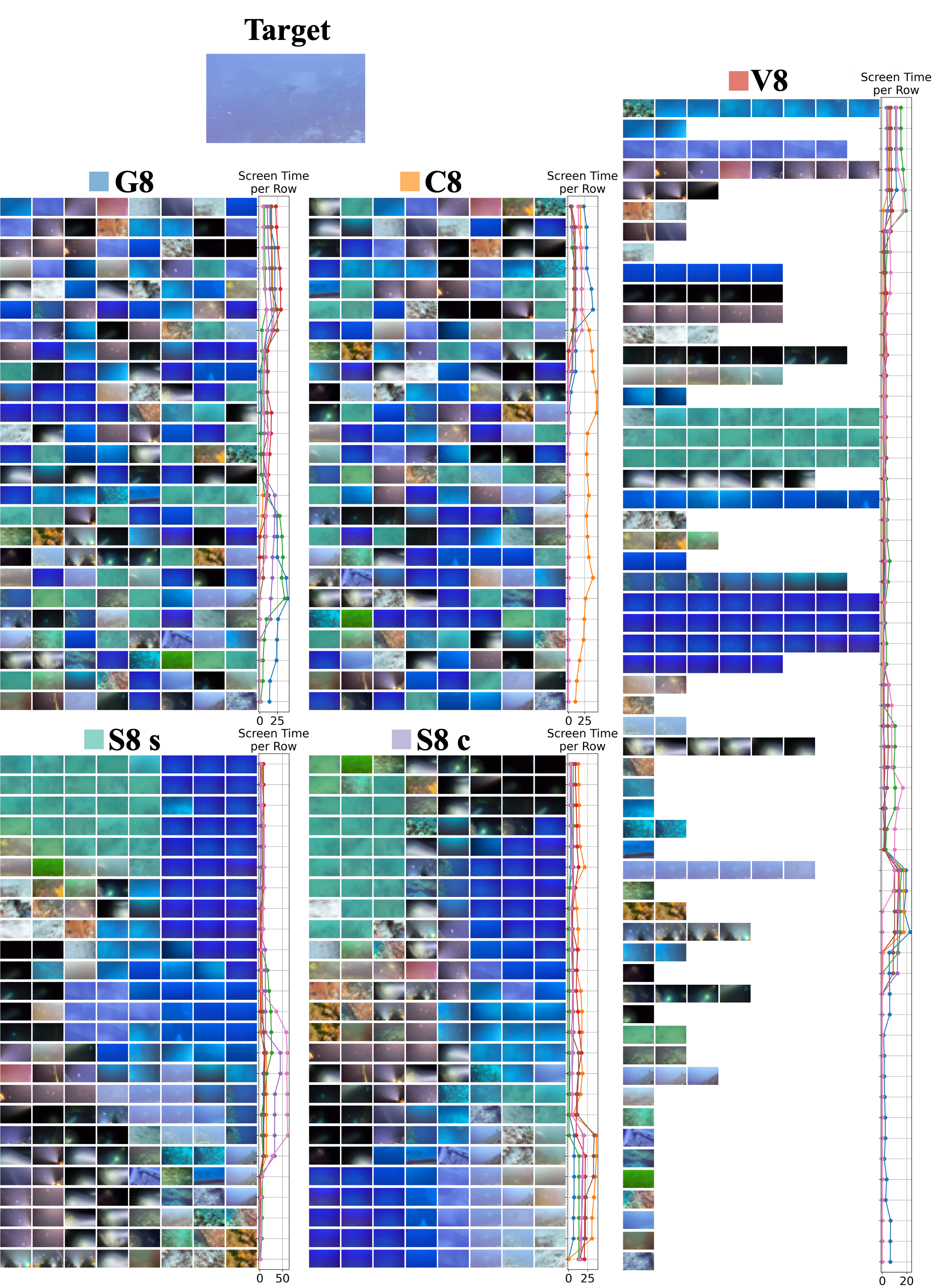}
    \caption{Target image at the top and eight-column layouts of 200 images with screen times of each row per user displayed on the right of each grid.}
    \Description{This figure presents five representative 8-column grids from different layouts, each paired with a line plot of screen time per row. For G8 and C8, viewing time is scattered across many rows, suggesting more uniform scanning. In contrast, the sorted and grouped layouts show a different pattern: rows that are dissimilar to the target receive very little viewing time and are often skipped entirely, while rows with high similarity concentrate the majority of the viewing time. The figure illustrates the “skipping rows” phenomenon.}
    \label{fig:qualitative_row_skipping}
\end{figure}

Previous work~\cite{JoosJKFPL24, ImageSearchStudy} found that users of retrieval systems rapidly navigate past entire regions of items with few or no fixations before reaching regions of interest. This behavior could also explain why participants were more effective for sorted layouts if the target was not in the collection: They ruled out uninteresting regions without explicitly focusing on them. We observed that for some collection-method pairs, most participants seemed to rule out entire rows quickly. We show an example in \autoref{fig:qualitative_row_skipping}: For the \GEight{} and \CEight{} layouts (top row), we observe that the screen time per row is scattered. Contrarily, for the sorted-based methods ( \TEightS{} and \TEightC{}) in the bottom row and \VEight{} on the right, screen times are focused on a few interesting regions where the candidates are visually similar to the target - blocks of entire rows are skimmed. Specifically, two participants ruled out entire rows in less than 0.5 seconds per row. Joos et al.~\cite{JoosJKFPL24} observed this behavior for ranked layouts too, if the candidate images are (coincidentally) grouped into similar regions with a low visual similarity to the target. We furthermore quantitatively measured the effect by counting how often eight or more rows were consecutively skipped, with each row having a screen time of two seconds or less. Note that on average, seven rows are displayed on the screen for 6 column layouts, meaning every row had an average screen time of at most $2s/7 \approx 285ms$. We measured 12 and 17 occurrences for \TEightC{} and two \TEightS{}, whereas we only measured 2 and 1 for \CEight{} and \GEight{}. 

The skipping-regions effect also help explain the notable differences in search times when the target is not part of the collection. In such cases, the sorted \TEightS{} layout significantly outperforms the four-grid layouts and is considerably faster than the other ranked-based approaches. \TEightC{} performs worse than its semantic counterpart, it still outperforms ranked layouts, although we measured significantly more scrolling. This result aligns with our earlier findings: when target positions are comparable, structured layouts tend to be more efficient. If the target is absent, participants must scan the entire grid - a task that benefits from a clear organized structure.
Moreover, since there is no target in the collection, the advantage of a low number of overlooks of four-column layouts is not applicable. This observation explains the inferior performance of 4-column layouts for collections without a target.

\begin{figure*}[h]
    \centering
    \includegraphics[width=\linewidth]{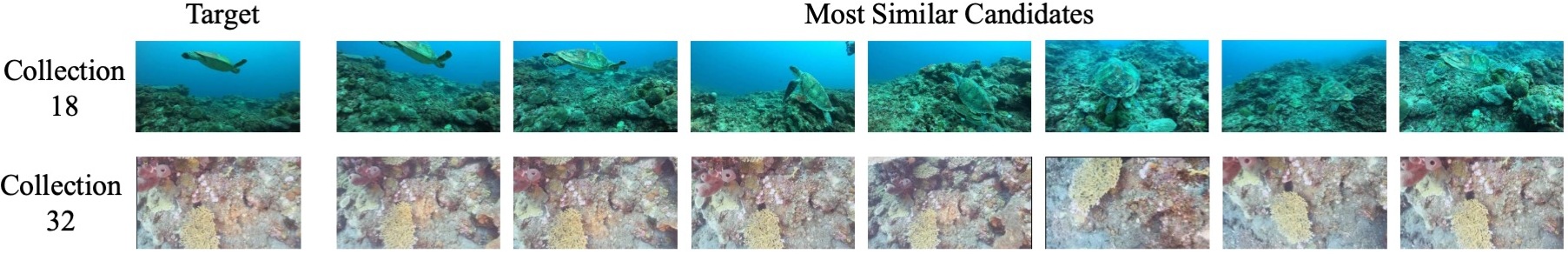}
    \Description{For two collections (18 and 32), the target image is shown with its seven most similar candidates. In collection 18, the target depicts a single turtle against light blue water, making it relatively easy to discriminate from candidates. In collection 32, the target shows several corals, producing more visual clutter, making it harder to discriminate. Although the target in collection 18 has a higher rank, all layouts perform better on collection 18. We hypothesize that target discriminability helps explain differences in search time.}
    \caption{Target and most similar candidates, measured with the cosine similarity and CLIP features, for collection 18 and 32.}
    \label{fig:qualitative_candidates}
\end{figure*}

\subsubsection{Collection Dependency} 
\begin{figure*}[!tb]
    \centering
    \includegraphics[width=\linewidth]{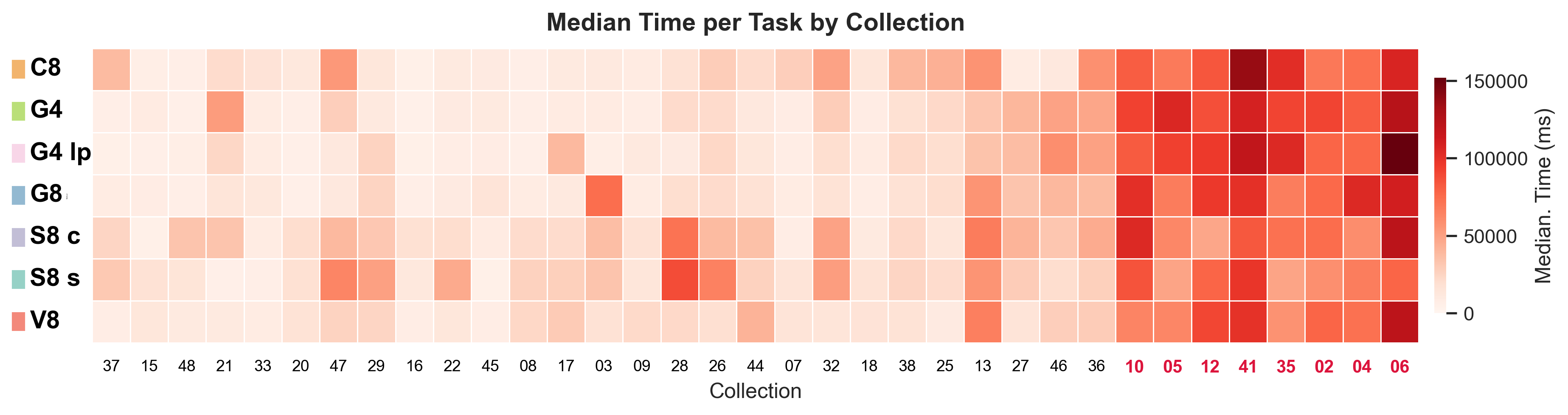}
    \Description{Heatmap showing median task time for every (layout, collection)-pair. Collections without a target yield longer search times. Rank-preserving layouts tend to be faster when the target has a good rank, while performance varies substantially across collections. Some collections (e.g., collection 13) are generally slow across all layouts, whereas most show a high layout-dependency. The heatmap demonstrates that multiple (counterbalancing) factors, e.g., target position, overlooks, and target discriminability, influence the search time, leading to a varying median search time for (layout, collection)-pairs.}
    \caption{Heatmap showing the median time participants spent on a task per method and collection. Collections are sorted based on their initial target item rank. Collections with red labels did not contain the target. We observe huge variations among collections. Especially collections without a target were time-consuming.}
    \label{fig:heatmap_average_time}
\end{figure*}

We emphasize that our reported results are averaged over a broad spectrum of collections in the homogeneous MVK dataset. However, \autoref{fig:heatmap_average_time} shows substantial variance among collections. While we identified the target rank as one determining factor, other factors, such as the specific image characteristics, also influence the search efficiency and behavior. For example, collection 18 has a higher target rank than collection 32, but participants were faster on collection 18 for all layouts. \autoref{fig:qualitative_candidates} reveals the difference between the collections: While collection 18 shows a single discriminative object (a turtle) that participants can focus on, collection 32 shows a sea floor with multiple less discriminative objects. We consider exploring which layout is preferable, depending on the visual characteristics of (target, candidates) pairs, an interesting topic for future work.

\subsubsection{Target Overlay and Comparison} Our study setup also offered the functionality to observe the target image again and to compare target images on a bigger screen with the candidate. While the overall average of the number of target uses is 1.23 and the average of comparisons is 2.76, there are some outliers with up to 28 comparisons. The average time spent on inspecting/comparing the target is 9.6 seconds, and thus makes up 22.4 \% of the overall search time. Although this amount is considerable, it is evenly distributed across the compared layouts and thus does not explain the observed differences.

\subsection{Effectiveness}\label{sec:accuracy}
Overall, two sources of errors can appear in our study setup: On the one hand, participants could submit a wrong candidate, and on the other hand, they could skip tasks, although the target image was in the collection. We report average values in \autoref{tab:overview}.

\subsubsection{Wrong Submissions} 
\begin{figure}
    \centering
    \includegraphics[width=1\linewidth]{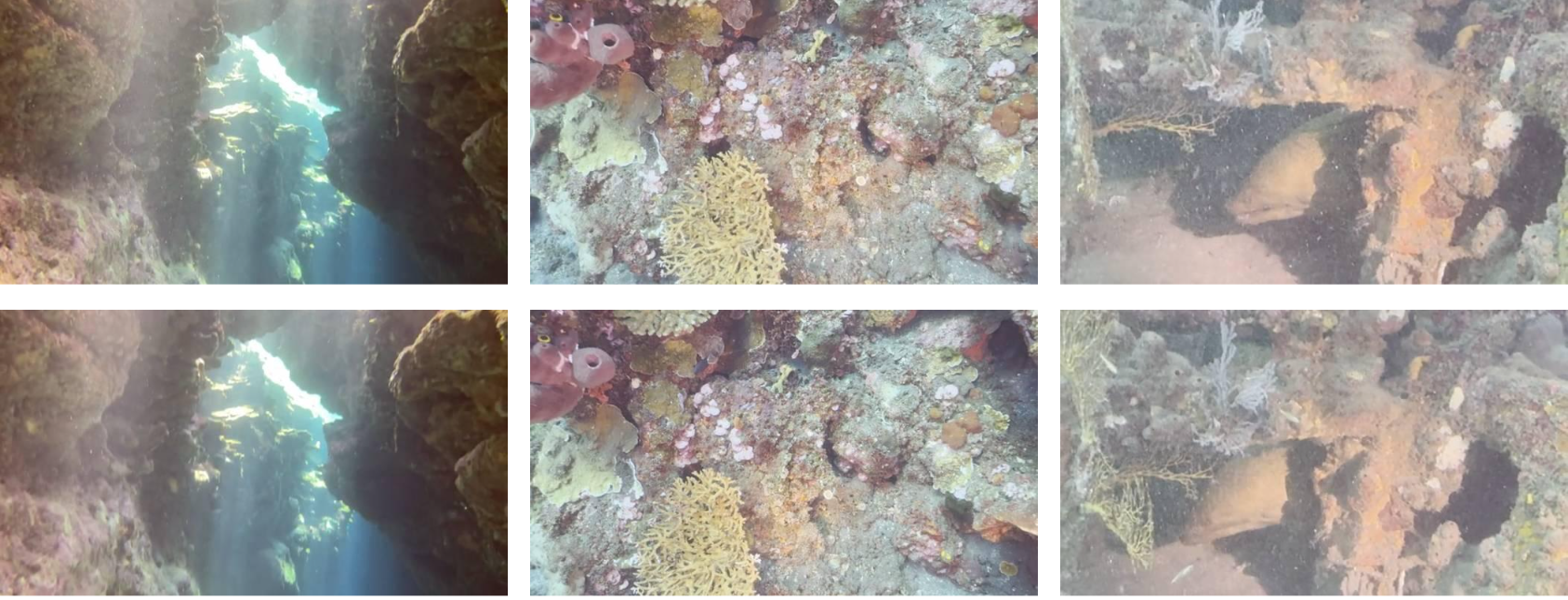}
    \Description{This figure shows three pairs of images; within each pair, the top image is the true target and the bottom is a close candidate. Together, the near-duplicate confusions of these three collections account for 95\% of wrong answers. In the first two pairs, the differences are tiny spatial shifts within otherwise identical scenes. In the third pair, a fish appears with a slightly opened mouth compared to the target.}
    \caption{Comparison of targets (top row) and corresponding wrong submission. The shown collections are 29, 32, and 36 (left to right), where 95\% of wrong submissions emerged.}
    \label{fig:wrong_submissions}
\end{figure}

We observed only rare cases of erroneous submissions. Across all datasets and layouts, we observe an average wrong submission rate of 5\%. This is expected, as our study setup allowed for a direct side-by-side comparison between candidates and the target. 95\% of errors occurred on only three collections. They contain visually close duplicates, as shown in \autoref{fig:wrong_submissions}. All mismatched frames are from the same video as the target, but not necessarily temporally close, with more than five seconds apart. We observe substantial differences in the error rate among the layouts. Especially sorted layouts suffer from a high error rate. We initially hypothesized that for the ranked-based layouts, the close duplicates are observed later than the target (e.g., at a later position in the grid). Consequently, the participants would not even observe the close duplicates. Contrarily, for sorted layouts, close candidates are usually positioned close to the targets and are observed at the same time. However, our hypothesis can not be confirmed: Even if close candidates are inspected first in the ranked and grouped layouts, they are not as often falsely submitted as in sorted layouts. Although future studies are necessary to confirm this effect, we suspect that participants get a false sense of security when spotting a \emph{cluster} of visually similar candidates to the target and subsequently submit the (seemingly) closest match.

In rare cases, the submitted candidates were visually different from the target. As participants qualitatively reported, this happened by accidental mouse clicks. Due to the large number of tasks that were solved without mistakes, we could not identify significant differences in the error rate between layouts based on pairwise Holm-Bonferroni corrected Wilcoxon Signed-Rank tests (see \autoref{supplementary_errors} for exact $p$-values).

\subsubsection{False Skips} From the 1323 iterations containing a target, only 57 tasks were not solved, leading to a false-skip rate of 4.3 \%. The 4-column layouts perform best, with participants falsely skipping four times less with the \GFourLP{} layout compared to sorted 8-column layouts. Although we do not observe statistically significant differences between methods based on a Holm-Bonferroni corrected Fisher test, we argue that the four-column layouts have clear advantages over all eight-column layouts. We report the exact $p$-values in \autoref{supplementary_false_skips}. The primary advantage is that overlooks do not occur as often, in line with our argumentation in \autoref{sec:efficiency}. In general, targets are overlooked almost twice as often with the sorted and video-grouped layouts compared to the 4-column layouts. Similarly, participants overlook the target 50\% more with the remaining 8-column layouts compared to the 4-column setting. In all falsely skipped cases, the target image was at least once on the participants' screens. Compared to wrong answers, false skips are more evenly distributed across collections. We show the exact distributions across layouts and collections in \autoref{appendix_collection_false_skips} and \autoref{appendix_collection_wrong_answers}.

%% file: Chapter/5-limitations.tex
We conducted a user study comparing seven keyframe layout strategies for the homogeneous MVK dataset, yielding several insights into when specific layout strategies are most beneficial and why. When targets were present in the candidate set, rank-based layouts were generally most effective, with the four-column grid with \GFourLP{} being the fastest. Contrarily, when targets were absent, participants benefited from sorted layouts such as \TEightS{}. Overall, the video grouped \VEight{} layout shows the fastest search time with the fewest erroneous submissions. By analyzing user interactions and the underlying keyframe collections, we found multiple behavioral factors explaining these findings. Firstly, sorted layouts suffer from position losses, where low-ranked targets get pushed further down the grid, delaying first-arrival times. They also suffer from more overlooks, further increasing search time and false skips. In contrast, 4-column ranked layouts produce the fewest overlooks. Thereby, we showed that not only the size of keyframes is crucial, but also the number of images displayed simultaneously on the screen. Contrary to previous eye-tracking studies, we found that these overlooks do not happen more in the horizontal center. Instead, we find that more overlooks appear in the vertical center. Furthermore, we detected an accumulation of overlooks on individual tasks, suggesting that the visual characteristics of the target and its neighborhood also highly influence overlooks. Finally, we found that sorted and grouped layouts profit from the efficient exclusion of whole blocks of images, counterbalancing the position losses and overlooks. An interesting area for future studies is to explore which visual characteristics lead to overlooks, and how to limit these in layouts.

The behavioral factors allow design implications beyond the MVK dataset. In scenarios where retrieval systems are expected to return the target among the top candidates, compact rank-based layouts should be preferred, as they minimize overlooks and keep the promising target positions. In contrast, when verifying that a collection does not contain the target or being confronted with higher target ranks, sorted or grouped layouts are advantageous, as they allow users to dismiss large unpromising regions quickly. A promising direction for future work is to combine layouts, e.g., by combining the \GFour{} layout for the first $k$ items, while the rest of the candidate set can be presented in an organized way using a sorted eight-column grid, such as \TEightS{}. The combined layout would profit from a low overlook rate of the most promising candidates, and the majority of low-ranked targets would still be positioned towards the top. At the same time, users can quickly scan over the subsequently sorted grid with potentially less relevant candidates. An alternative is a rank-aware sorting that inherently organizes items based on their similarity and rank as demonstrated in rating-aware self-organizing maps~\cite{PeskaL22}. 

Despite careful consideration of the setup for our user evaluation, the work comes with some limitations that we discuss below: Firstly, our study is only a comparative starting point for efficient video retrieval.
Due to the extensive related work, we could only investigate a sample of possible layouts.
Thereby, we focused on the layouts most used in competitive Video Retrieval Systems where time and accuracy matter. Furthermore, we focused our study on the performance of these layouts at the expense of user experience. However, users might feel more confident using different layouts, which would also explain the higher error rate of sorted grids, as hypothesized in \autoref{sec:accuracy}.

A second limitation is that our findings are related to the underlying dataset. The MVK dataset is highly homogeneous, often containing keyframes from a single video. This strongly affects the performance of multiple layouts: For example, \VEight{} degenerates if the candidates are from diverse videos. Thus, a future work is to explore alternatives to \VEight{}, which groups keyframes of video, while not wasting screen space. One could formalize this as an optimization problem, where the distance of keyframes from one video is minimized, while the occupied screen space is maximized.
Furthermore, we found that sorting methods profit from participants excluding large regions of rows.
However, more heterogeneous datasets contain more heterogeneous colors. It is thus unclear if the effects transfer to those datasets.

Furthermore, our study is limited to the visual KIS task. For other tasks, such as textual Known-Item Search, users pay less attention to colors than to semantic information. Thus, it is unclear how our findings transfer to other tasks. We consider a similar study, with different tasks, as a valuable direction for future work. Moreover, we assumed in our setup that users have a clear visual representation of what they want to retrieve. Thus, we allowed participants to observe the target as often as they wanted to and additionally provided a side-by-side comparison. Search behavior might also change if users do not have a clear imagination of visual attributes, e.g., shades of a color. As the displayed target represents a perfect memory recall, we expect the times to be rather lower bounds of real recall times. However, a thorough investigation of this phenomenon remains beyond the scope of this work. The overall reported search time consists of the actual browsing time and time spent inspecting the target using the target and comparison views. \autoref{tab:overview} does not show large differences in these view form times, and therefore, we hypothesize that these times are affected more by user memorization skills and task complexity than layout design.

%\begin{itemize}
%    \item The study focused on visual known item search with full information about target, which is a specific search scenario that differs from other search scenarios. The same set of layouts can be compared with other task types (e.g., textual known-item search, or ad-hoc search) with potentially different outcomes. We also note that the time is affected by the necessity to find one particular keyframe. The times and outcomes may change if any keyframe from the whole shot or video is acceptable as the result of searching. We leave this experiment to future work.
%    \item The age of participants and screen sizes have a non-trivial effect on study outcomes and thus we note that more studies are needed. In other words, the presented findings do not have to generalize to older users or to significantly smaller/larger displays, where some layouts might even not be appropriate. 
%    \item The overall reported search time consists of the actual search time and time spent with ``remembering'' target using target and comparison views. \autoref{tab:overview} does not show large differences in these view form times, and therefore we hypothesize that these times are affected more by user memorization skills and task complexity than layout design. As the displayed target represents a perfect memory recalling, we expect the times to be rather lower bounds of real recall times. Understanding this phenomenon is definitely beyond the scope of this paper. 

%\end{itemize}

%% file: Chapter/6-conclusion.tex
This work presented the first large-scale comparative study of keyframe layouts for Visual Known-Item Search. Across 49 participants and 1715 tasks, we systematically evaluated seven commonly used layouts in terms of search efficiency and accuracy. We found statistically significant search-time differences when conditioning on whether the target was present: when the target was in the collection, rank-preserving layouts were advantageous; when it was absent, sorted layouts were faster for confirming non-existence. Overall, a video-grouped layout achieved the lowest mean search time. For accuracy, four-column ranked grids tended to reduce overlooks and false skips relative to other layouts, while sorted grids saw more wrong submissions. We provide evidence that organized structures are suitable for quickly scanning entire regions of homogeneous regions, but suffer from higher overlooks and also push promising candidates further down the grid, leading to higher search times. These findings argue against a one-size-fits-all layout and instead suggest hybrids that preserve rank for top-k candidates while enabling efficient browsing of the remainder, either by combining ranked and sorted views or by employing rank-aware sorting. Our study used a homogeneous underwater dataset and a visual KIS setup with target re-inspection; extending to heterogeneous datasets and other retrieval tasks remains future work.

%% file: Chapter/7-appendix.tex
\section{Hyperparameters for Sorting Algorithms}\label{sec:appendix_hyperparameters}
In the following, we provide details necessary to reproduce the \TEightC{} and \TEightS{} layouts. Both algorithms rely on image-level features. Thus, we extracted color features for the \TEightC{} algorithm, and semantic CLIP-based features for \TEightS{}. Specifically, we extracted for each keyframe the average values of each color channel in the LAB color space for \TEightC{}. For \TEightS{}, we extracted embeddings from a CLIP model optimized for image-image queries~\cite{SchallBHJ24} that is also applied in VBS~\cite{stroh2025prak}.

For sorting, we use the FLAS algorithm~\cite{flas}, available at GitHub\footnote {https://github.com/Visual-Computing/LAS\_FLAS}. FLAS arranges keyframes on the grid by starting from a random one-to-one placement and then iteratively improving the layout with local swaps within a shrinking neighborhood. Thereby, FLAS relies on three hyperparameters:
\begin{itemize}
\item The initial filter radius, which we set to 0.5
\item The radius reduction factor, which we set to 0.7
\item The number of swap candidates, which we set to 49.
\end{itemize}
We found that these hyperparameters lead to satisfactory sorting results.

\section{Rank Distribution}\label{supplementary_rank_pos}
\begin{figure}[H]
    \centering
    \includegraphics[width=\linewidth]{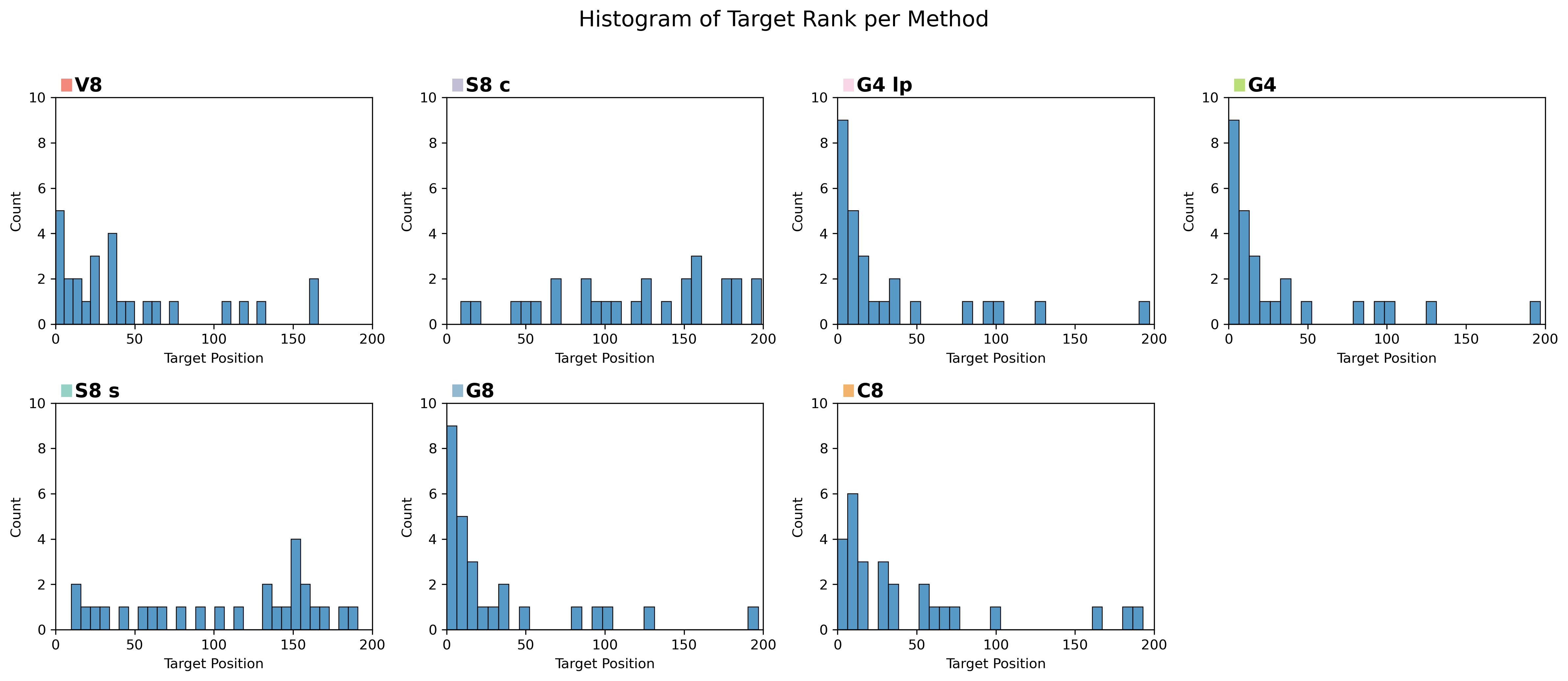}
    \Description{Seven bar charts showing where targets appear in row-major order for each of the seven layouts. The ranked layouts (e.g., G4, G4 with side panel, G8, C8) are left-heavy, meaning targets frequently occupy early positions. The sorted layouts (S8s, S8c) show a more uniform position distribution, reflecting that similarity-based grouping decouples position from rank. The video-grouped layout also skews toward early positions, but overall pushes some targets downward.}
    \caption{Distribution of flattened target positions per layout. The target position is the row-major flattened representation of the grid.}
    \label{fig:rank_distribution}
\end{figure}

We show the target positions in \autoref{fig:rank_distribution}. From the positions of the rank-based layouts (e.g. \GFour{}), we see an effective ranking model with predominantly low target ranks. We emphasize that this target distribution is realistic for MVK, directly resulting from text queries. The text queries stem from two annotators experienced in video retrieval, but without knowledge of the underwater domain. Due to the overall low target ranks, sorting algorithms suffer from a position loss of the targets, resulting in an increase in their first arrival time.

\section{Supplementary Experimental Evaluation}
We show further experimental results in the following subsections, mentioned in the main text. 

\subsection{Significant Differences between Layouts of False Skips}\label{supplementary_false_skips}
We evaluate whether there are significant differences in False Skips between layouts using Fisher’s test with Holm-Bonferroni Corrections. The results are shown in \autoref{tab:false_skips}.

\begin{table}[ht]
\centering

\caption{Holm-Bonferroni corrected $p$-values for the number of False Skips (upper triangle) and the respective uncorrected  $p$-values (lower triangle).}
\renewcommand{\arraystretch}{1.2} % etwas mehr Zeilenhöhe
\begin{tabular}{
    l|
    c c c c c c c
}
\toprule
\textbf{Row} & 
\GFour{} & 
\GFourLP{} & 
\GEight{} & 
\CEight{} & 
\TEightC{} & 
\TEightS{}& 
\VEight{} \\
\midrule
  \GFour{} & -  & 1  & 1  & 1  & 1  & 1  & 1 \\ \GFourLP{} & 0.338  & -  & 1  & 1  & 0.666  & 1  & 1 \\ \GEight{} & 1  & 0.221  & -  & 1  & 1  & 1  & 1 \\ \CEight{} & 0.471  & 0.054  & 0.640  & -  & 1  & 1  & 1 \\ \TEightC{} & 0.348  & \textbf{0.033}  & 0.493  & 1  & -  & 1  & 1 \\ \TEightS{} & 0.800  & 0.141  & 1  & 0.820  & 0.655  & -  & 1 \\ \VEight{} & 1  & 0.338  & 1  & 0.471  & 0.348  & 0.800  & - \\
\bottomrule
\end{tabular}
\label{tab:false_skips}
\end{table}

\subsection{Collection Dependency of False Skips}\label{appendix_collection_false_skips}
\begin{figure}[H]
    \centering
    \includegraphics[width=1\linewidth]{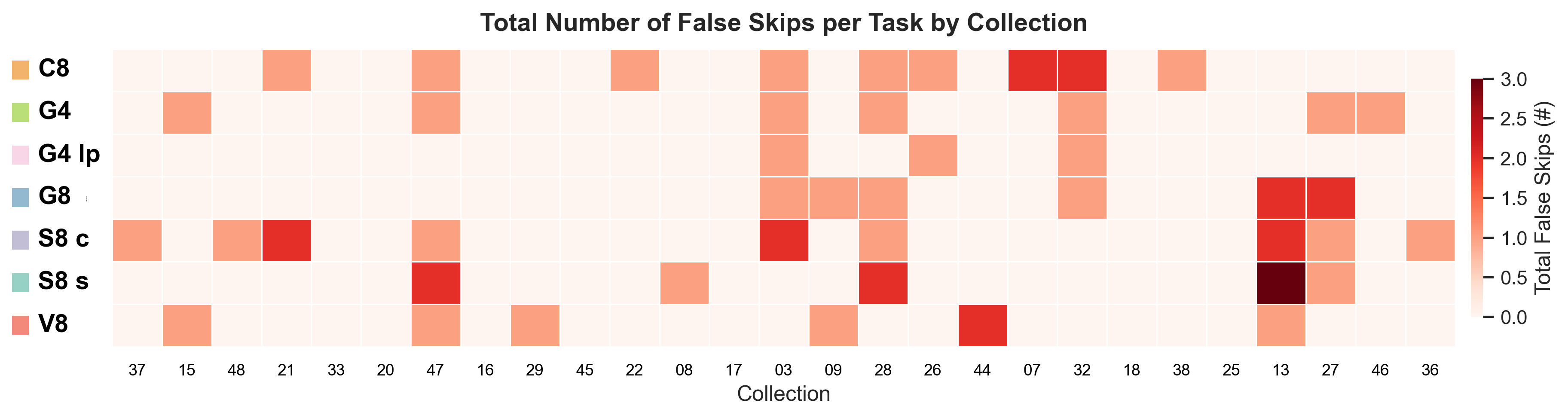}
    \Description{Heatmap showing the total number of False Skips for every (layout, collection) pair. While some collections (especially collections 13 and 27) suffer from a higher number of false skips (with a maximum of 3 for a (layout, collection) pair, false skips happened on almost all collections. Only 7 out of the 28 collections with a target showed no false skip.}
    \caption{Heatmap showing the total number of False Skips on a task per method and collection. Collections are sorted based on their initial target item rank.}
    \label{fig:avg_falseskips_task}
\end{figure}

We show the total number of False Skips per (collection, layout) pair in \autoref{fig:avg_falseskips_task}. False Skips are more evenly distributed across collections compared to erroneous answers.

\subsection{Significant Differences between Layouts of Overlooks}\label{supplementary_overlooks}
We evaluate whether there are significant differences in Overlooks between layouts using Wilcoxon's Signed Rank test with Holm-Bonferroni Correction. The results are shown in \autoref{tab:overlooks}.

\begin{table}[H]
\caption{Holm-Bonferroni corrected $p$-values for the number of Overlooks (upper triangle) and the respective uncorrected  $p$-values (lower triangle).}
\centering
\renewcommand{\arraystretch}{1.2} % etwas mehr Zeilenhöhe
\begin{tabular}{
    l|
    c c c c c c c
}
\toprule
\textbf{Row} & 
\GFour{} & 
\GFourLP{} & 
\GEight{} & 
\CEight{} & 
\TEightC{} & 
\TEightS{}& 
\VEight{} \\
\midrule
 \GFour{} & -  & 1  & 1  & 1  & 1  & 0.537  & 0.287 \\ \GFourLP{} & 0.967  & -  & 1  & 1  & 0.883  & 0.577  & 0.297 \\ \GEight{} & 0.248  & 0.230  & -  & 1  & 1  & 1  & 1 \\ \CEight{} & 0.079  & 0.079  & 0.474  & -  & 1  & 1  & 1 \\ \TEightC{} & 0.086  & 0.053  & 0.354  & 0.850  & -  & 1  & 1 \\ \TEightS{} & \textbf{0.029}  & \textbf{0.033}  & 0.244  & 0.641  & 0.567  & -  & 1 \\ \VEight{} & \textbf{0.015}  & \textbf{0.016}  & 0.294  & 0.710  & 0.607  & 0.858  & - \\
\bottomrule
\end{tabular}
\label{tab:overlooks}
\end{table}

\subsection{Significant Differences between Layouts of Erroneous Answers}\label{supplementary_errors}
We evaluate whether there are significant differences in erroneous answers between layouts using Wilcoxon's Signed Rank test with Holm-Bonferroni Correction. The results are shown in \autoref{tab:errors}. Overall, we find a higher number of erroneous answers for sorted layouts, but do not find statistical significant differences.

\begin{table}[H]
\centering
\renewcommand{\arraystretch}{1.2}
\setlength{\tabcolsep}{5pt}
\caption{$p$-values for pairwise Wilcoxon Signed-Rank tests for the number of wrongly submitted answers. The lower triangle represents the uncorrected $p$-values, while the upper triangle shows the Holm-Bonferroni $p$-values.}
% ---- BLOCK 1: Total ----
\begin{tabular}{l|ccccccc}

\toprule
\multicolumn{8}{c}{\textbf{Total}}\\
\midrule
      & \GFour{} & \GFourLP{} & \GEight{} & \CEight{} & \TEightC{} & \TEightS{} & \VEight{} \\
\midrule
\GFour{}   & - & 1.000 & 1.000 & 1.000 & 0.950 & 1.000 & 1.000 \\
\GFourLP{} & 1.000 & - & 1.000 & 1.000 & 0.950 & 1.000 & 1.000 \\
\GEight{}  & 0.637 & 0.637 & - & 1.000 & 0.403 & 1.000 & 1.000 \\
\CEight{}  & 0.835 & 0.819 & 0.491 & - & 1.000 & 1.000 & 1.000 \\
\TEightC{} & 0.050 & 0.050 & \textbf{0.019} & 0.076 & - & 1.000 & 0.403 \\
\TEightS{} & 0.356 & 0.239 & 0.117 & 0.317 & 0.347 & - & 1.000 \\
\VEight{}  & 0.655 & 0.637 & 1.000 & 0.467 & \textbf{0.019} & 0.102 & - \\
\bottomrule
\end{tabular}

\vspace{0.6em}

% ---- BLOCK 2: Target in Collection ----
\begin{tabular}{l|ccccccc}
\toprule
\multicolumn{8}{c}{\textbf{Target in Collection}}\\
\midrule
      & \GFour{} & \GFourLP{} & \GEight{} & \CEight{} & \TEightC{} & \TEightS{} & \VEight{} \\
\midrule
\GFour{}   & - & 1.000 & 1.000 & 1.000 & 1.000 & 1.000 & 1.000 \\
\GFourLP{} & 0.819 & - & 1.000 & 1.000 & 0.422 & 1.000 & 1.000 \\
\GEight{}  & 0.317 & 0.405 & - & 1.000 & 0.127 & 1.000 & 1.000 \\
\CEight{}  & 0.655 & 0.796 & 0.593 & - & 0.365 & 1.000 & 1.000 \\
\TEightC{} & 0.064 & \textbf{0.023} & \textbf{0.006} & \textbf{0.019} & - & 1.000 & 0.127 \\
\TEightS{} & 0.349 & 0.221 & 0.061 & 0.162 & 0.274 & - & 1.000 \\
\VEight{}  & 0.317 & 0.439 & 1.000 & 0.593 & \textbf{0.006} & 0.061 & - \\
\bottomrule
\end{tabular}

\vspace{0.6em}

% ---- BLOCK 3: Target not in Collection ----
\begin{tabular}{l|ccccccc}
\toprule
\multicolumn{8}{c}{\textbf{Target not in Collection}}\\
\midrule
      & \GFour{} & \GFourLP{} & \GEight{} & \CEight{} & \TEightC{} & \TEightS{} & \VEight{} \\
\midrule
\GFour{}   & - & 1.000 & 1.000 & 1.000 & \textless{}0.001 & 1.000 & 1.000 \\
\GFourLP{} & 0.317 & - & 1.000 & 1.000 & 1.000 & 1.000 & 1.000 \\
\GEight{}  & 0.157 & 0.564 & - & 1.000 & 1.000 & 1.000 & 1.000 \\
\CEight{}  & 0.083 & 0.157 & 0.564 & - & 1.000 & 1.000 & 1.000 \\
\TEightC{} & \textless{}0.001 & 0.317 & 0.157 & 0.083 & - & 1.000 & 1.000 \\
\TEightS{} & 0.317 & 1.000 & 0.564 & 0.317 & 0.317 & - & 1.000 \\
\VEight{}  & 0.157 & 0.564 & 1.000 & 0.655 & 0.157 & 0.317 & - \\
\bottomrule
\end{tabular}
\label{tab:errors}
\end{table}

\subsection{Collection Dependency of Erroneous Answers}\label{appendix_collection_wrong_answers}
\begin{figure}[H]
    \centering
    \includegraphics[width=1\linewidth]{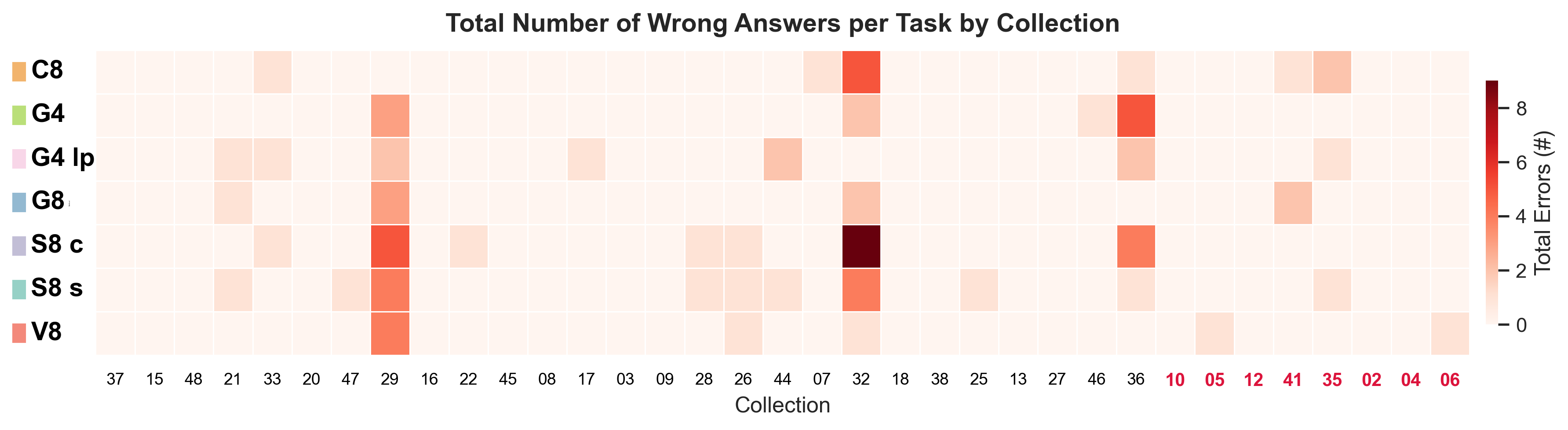}
    \Description{Heatmap showing the total number of Erroneous Answers for every (layout, collection) pair. Almost all collections show a low error rate, except for three collections (29,32,36). Especially collections 29 and 32 show a high number of Erroneous Answers for almost all (collection, layout) pairs, with a maximum value of 8.}
    \caption{Heatmap showing the total number of Erroneous Answers on a task per method and collection. Collections are sorted based on their initial target item rank.}
    \label{fig:avg_wrong_task}
\end{figure}

We show the total number of Erroneous Answers per (collection, layout) pair in \autoref{fig:avg_wrong_task}. Erroneous Answers are highly concentrated on only three layouts, resulting from close candidates (see \autoref{fig:wrong_submissions}.

\section{Screen Size}\label{appendix_screen_size}
21 participants conducted the study using a 27-inch Dell UltraSharp U2722DE display, while 28 used a 24-inch AOC e2460Sh. Both monitors used a 16:9 aspect ratio. Consequently, both configurations presented the same content within the viewport (e.g., the same number of rows). Time per task was slightly higher with a 27-inch monitor ($n = 735$, mean$= 45.5$) than with a 24-inch monitor ($n = 980$, median$= 40.5$). We further emphasize that the search time between participants varies strongly. The standard deviation of per-user average search times is 10.7 seconds. However, a two-sided Mann–Whitney U test indicated statistical significance with a $p$-value of $.018$ and Cliff's $\delta = 0.066$, suggesting a slight tendency for longer times with the 27-inch monitor. Using Mann-Whitney U tests, we do not find statistically significant differences in the number of overlooks between screens ($p=0.397$), nor in the number of wrong answers ($p=0.757$). We further report on the per-layout search times in \autoref{tab:overview_by_screen_grouped}. While we observe some per-screen deviations, specifically in search times and false skips of \GEight{}, our general findings are confirmed across all layouts: Four-column layouts produce fewer overlooks than eight-column layouts, and especially sorted layouts suffer from overlooks. Furthermore, ranked layouts are more efficient if the target is contained in the collection, while sorted and grouped layouts are suitable to exclude the existence of targets. 
% Preamble (ensure these are loaded)
% \usepackage{booktabs}
% \usepackage{array}
% \usepackage{multirow}

\begin{table*}[h]
\caption{We report averaged values over all tasks for four measurements, partitioned into screen size. \underline{Time} is the overall search time from task start to submit/skip, \underline{First Arr.} is the first-arrival time (first time target appeared on the screen), \underline{F. Skips} is the number of false skips, \underline{Overlooks} is the number of overlooks of the target}
\centering
\renewcommand{\arraystretch}{1.2}
\setlength{\tabcolsep}{5pt}

% 1 label column + 4 groups × (24",27") = 9 columns total
\begin{tabular}{l|cc!{\vrule width 1.2pt}cc!{\vrule width 1.2pt}cc!{\vrule width 1.2pt}cc}
\toprule
& \multicolumn{2}{c}{\textbf{Time (s) $\downarrow$}}
& \multicolumn{2}{c}{\textbf{First Arr. (s) $\downarrow$}}
& \multicolumn{2}{c}{\textbf{F. Skips (\%) $\downarrow$}}
& \multicolumn{2}{c}{\textbf{Overlooks (\%) $\downarrow$}} \\
\cmidrule(lr){2-3}\cmidrule(lr){4-5}\cmidrule(lr){6-7}\cmidrule(lr){8-9}
\textbf{Layout} & \textbf{24-inch} & \textbf{27-inch} & \textbf{24-inch} & \textbf{27-inch} & \textbf{24-inch} & \textbf{27-inch} & \textbf{24-inch} & \textbf{27-inch} \\
\midrule

\multicolumn{9}{l}{\textbf{Total}} \\
\GFour{}   & 41.0 & 45.0 & 7.1 & 9.2 & 2.8 & 4.9 & 19.4 & 21.0 \\
\GFourLP{} & 40.1 & 45.3 & 7.3 & 7.2 & 1.9 & 1.2 & 22.2 & 19.8 \\
\GEight{}  & 38.8 & 47.4 & 6.8 & 4.1 & 1.9 & 7.4 & 24.1 & 32.1 \\
\CEight{}  & 42.1 & 46.4 & 3.0 & 5.5 & 4.6 & 7.4 & 25.9 & 43.2 \\
\TEightC{} & 43.0 & 46.3 & 12.4 & 12.9 & 6.5 & 6.2 & 31.5 & 38.3 \\
\TEightS{} & 38.9 & 47.6 & 9.6 & 10.0 & 3.7 & 6.2 & 33.3 & 50.6 \\
\VEight{}  & 39.7 & 40.6 & 4.8 & 3.8 & 3.7 & 3.7 & 39.8 & 29.6 \\
\midrule

\multicolumn{9}{l}{\textbf{Target in Collection}} \\
\GFour{}   & 23.6 & 27.1 & 7.1 & 9.2 & 2.8 & 4.9 & 19.4 & 21.0 \\
\GFourLP{} & 22.9 & 26.2 & 7.3 & 7.2 & 1.9 & 1.2 & 22.2 & 19.8 \\
\GEight{}  & 22.6 & 28.3 & 6.8 & 4.1 & 1.9 & 7.4 & 24.1 & 32.1 \\
\CEight{}  & 27.8 & 30.5 & 3.0 & 5.5 & 4.6 & 7.4 & 25.9 & 43.2 \\
\TEightC{} & 31.3 & 33.7 & 12.4 & 12.9 & 6.5 & 6.2 & 31.5 & 38.3 \\
\TEightS{} & 28.1 & 37.1 & 9.6 & 10.0 & 3.7 & 6.2 & 33.3 & 50.6 \\
\VEight{}  & 25.8 & 24.2 & 4.8 & 3.8 & 3.7 & 3.7 & 39.8 & 29.6 \\
\midrule

\multicolumn{9}{l}{\textbf{Target not in Collection}} \\
\GFour{}   & 99.7 & 105.5 & - & - & - & - & - & - \\
\GFourLP{} & 98.1 & 109.8 & - & - & - & - & - & - \\
\GEight{}  & 93.5 & 112.1 & - & - & - & - & - & - \\
\CEight{}  & 90.6 & 99.9  & - & - & - & - & - & - \\
\TEightC{} & 82.3 & 88.8  & - & - & - & - & - & - \\
\TEightS{} & 75.4 & 83.2  & - & - & - & - & - & - \\
\VEight{}  & 86.5 & 96.1  & - & - & - & - & - & - \\
\bottomrule
\end{tabular}
\label{tab:overview_by_screen_grouped}
\end{table*}